\documentclass{imamat}

\usepackage{color}
\usepackage{amsmath}
\usepackage{graphicx}
\usepackage{array}
\usepackage{url} 

\begin{document}

\title{Origin, bifurcation structure and stability of localized states in Kerr dispersive optical cavities}

\shorttitle{Localized structures in Kerr dispersive cavities} 
\shortauthorlist{P. Parra-Rivas, E. Knobloch, L. Gelens, and D. Gomila} 

\author{
\name{P. Parra-Rivas$^*$}
\address{OPERA-photonics, Universit\'e libre de Bruxelles, 50 Avenue F. D. Roosevelt, CP 194/5, B-1050 Bruxelles, Belgium\email{$^*$pparrari@ulb.ac.be}}
\name{E. Knobloch}
\address{Department of Physics, University of California, Berkeley CA 94720, USA}
\name{L. Gelens}
\address{Laboratory of Dynamics in Biological Systems, KU Leuven Department of Cellular and Molecular Medicine, University of Leuven, B-3000 Leuven, Belgium}
\and
\name{D. Gomila}
\address{Instituto de F\'{\i}sica Interdisciplinar y Sistemas Complejos, IFISC (CSIC-UIB), Campus Universitat de les Illes
	Balears, E-07122 Palma de Mallorca, Spain}}

\maketitle

\begin{abstract}
{Localized coherent structures can form in externally-driven dispersive optical cavities with a Kerr-type nonlinearity. Such systems are described by the Lugiato-Lefever equation, which supports a large variety of dynamical solutions. Here, we review our current knowledge on the formation, stability and bifurcation structure of localized structures in the one-dimensional Lugiato-Lefever equation. We do so by focusing on two main regimes of operation: anomalous and normal second-order dispersion. In the anomalous regime, localized patterns are organized in a homoclinic snaking scenario, which is eventually destroyed, leading to a foliated snaking bifurcation structure. In the normal regime, however, localized structures undergo a different type of bifurcation structure, known as collapsed snaking.}
{Bifurcation structure, homoclinic snaking, collapsed snaking, nonlinear optics}
\\
2000 Math Subject Classification: 34K30, 35K57, 35Q80,  92D25
\end{abstract}

\section{Introduction}

Localized dissipative structures, hereafter referred to as LSs, emerge in a vast variety of systems out of thermodynamic equilibrium, ranging from plasma physics and plant ecology to nonlinear optics and biology \citep{akhmediev_dissipative_2008,descalzi_localized_2011}. The formation of such states is associated with a double balance, between nonlinearity and spatial coupling (e.g., diffusion, dispersion and/or diffraction) on the one hand, and energy dissipation and gain or driving on the other \citep{akhmediev_dissipative_2008}. Therefore, the formation of LSs is not related to the presence of intrinsic inhomogeneities in the system. In nonlinear optics, the confinement of light in optical cavities may lead to the formation of LSs, which can be stationary or exhibit spatio-temporal dynamical behavior including oscillations, excitability, and chaos \citep{descalzi_localized_2011}. In these cavities, the role of spatial coupling is played either by beam diffraction or chromatic dispersion. In the first case, LSs have been studied in externally-driven diffractive nonlinear Kerr cavities \citep{scroggie_pattern_1994,firth_two-dimensional_1996,firth_dynamical_2002}. In this case, LSs consist of two-dimensional spots of light embedded in a homogeneous background, and they form in the plane transverse to the light propagation direction. These LSs are therefore commonly known as {\it spatial cavity solitons}. In the second case, one-dimensional LSs form in wave-guided dispersive Kerr cavities, such as fiber cavities, whispering gallery mode resonators, and micro- resonators, where the localization takes place along the propagation direction. In this context, LSs are typically called {\it temporal cavity solitons}. Temporal LSs were experimentally demonstrated for the first time by \citet{leo_temporal_2010} in the context of passive fiber cavities, and they were proposed as key elements for all-optical information buffering. After this initial observation, the interest in temporal LSs has grown rapidly, in part due to their application in frequency comb generation \citep{delhaye_optical_2007,kippenberg_microresonator-based_2011}, which led in turn to the discovery of a wide range of different types of stationary and dynamical LSs \citep{leo_dynamics_2013,herr_temporal_2014-1,xue_mode-locked_2015,garbin_experimental_2017}.

Here, we review the origin, bifurcation structure and stability of the different types of temporal LSs arising in passive Kerr dispersive cavities in both anomalous and normal dispersion regimes. In the mean-field approximation, such cavities can be modeled by the well-known Lugiato-Lefever (LL) equation \citep{lugiato_spatial_1987,chembo_theory_2017} 
\begin{align}\label{LLE}
\partial_t A=-(1+i\Delta)A+i\nu \partial_x^2A+i|A|^2A+S,
\end{align}
with periodic boundary conditions
\begin{equation}
A(L+x,t)=A(x,t), \hspace{0.5cm} \partial_xA(L+x,t)=\partial_xA(x,t)
\end{equation}
corresponding to a periodic domain of (large) period $L$. Here $A$ represents the normalized slowly varying amplitude of the electric field circulating in the cavity, $\Delta$ is the normalized intra-cavity phase detuning, and $S>0$ is the normalized driving field amplitude or pump. The parameter $\nu=\pm1$, with $\nu=1$ for the anomalous dispersion regime and $\nu=-1$ for the normal regime. In the following, we focus on solutions of Eq.~(\ref{LLE}) that respect the reflection symmetry $x\rightarrow-x$ of the equation, but also study solutions that break this symmetry. We use $\Delta$ and $S$ as control parameters, once $\nu$ is fixed, and take $L=160$, solving Eq.~(\ref{LLE}) on the domain $-80\le x\le 80$.

In optics, the LL equation was first derived in the context of passive diffractive Kerr cavities \citep{lugiato_spatial_1987}, and later on it was used to describe dispersive Kerr cavities, such as fiber cavities \citep{haelterman_dissipative_1992}, microresonators \citep{coen_modeling_2013}, and whispering gallery mode resonators \citep{chembo_spatiotemporal_2013}. However, the LL equation had in fact appeared even earlier in the context of plasma physics and condensed matter physics \citep{morales_ponderomotive-force_1974,kaup_theory_1978}.

In one spatial dimension, the appearance of LSs is usually related to the presence of bistability between two different, but coexisting states, and their formation is mediated by the locking or pinning of fronts or domain walls (DWs) corresponding to heteroclinic orbits in a spatial dynamics description of the system. One plausible situation is that an homogeneous state coexists with a subcritical Turing pattern \citep{tlidi_1994}. In this case, the locking of the DWs between such states leads to the formation of LSs consisting of a slug of the pattern embedded in a homogeneous background. Such structures are known as {\it localized patterns} (LPs). In the context of the LL equation [Eq.~(\ref{LLE})], this scenario appears in the anomalous dispersion regime, where LSs arise in the form of {\it bright} LPs. A second situation is related to the presence of bistability between two different homogeneous states. A LS can then be seen as a portion of one homogeneous state embedded in the other \citep{coullet_localized_2002}. This is the scenario that one encounters in the normal dispersion regime, where the typical LSs are {\it dark}.

Owing to different DW locking processes, the LSs exhibit bifurcation structures with distinct morphologies. In the anomalous regime, LPs undergo a {\it snakes-and-ladders} structure, whose main skeleton consists of two intertwined LP solution curves, which oscillate back-and-forth within a well-defined parameter range, while increasing the LP width. Due to its particular shape, this bifurcation structure is referred to as {\it homoclinic snaking}. The concept of homoclinic snaking goes back to the late 90s and the seminal paper {\it Heteroclinic tangles and homoclinic snaking in the unfolding of a degenerate reversible Hamiltonian Hopf bifurcation} where Woods and Champneys explain the formation of LPs through geometrical considerations, laying the foundations of an important new field of study \citep{woods_heteroclinic_1999}. In the normal regime, however, the dark LSs are organized differently, in the so-called {\it collapsed homoclinic snaking} structure, a structure that is related to the presence of oscillatory tails in the DW profiles \citep{knobloch_homoclinic_2005}. Our main concern in this article is to provide a detailed discussion of these two different bifurcation structures in the context of passive dispersive Kerr cavities. To do so we review the most relevant studies regarding this topic, before presenting in Sec.~\ref{sec:22} new results that are essential to understanding the emergence of these scenarios.

The paper is organized as follows. In Sec.~\ref{sec:1} we introduce the stationary problem, and start by studying the homogeneous steady states and their linear stability properties in the main regimes of operation (Sec.~\ref{sec:1.1}). We also present a spatial dynamical analysis of the system, where we identify the bifurcations from which LSs may emerge, and classify the different equilibria of the equation (Sec.~\ref{sec:1.2}). Next, in Sec.~\ref{sec:2}, we use multiscale perturbation methods to reduce Eq.~(\ref{LLE}) to different normal forms around each of the previously identified bifurcations, and use these to find time-independent small amplitude LS solutions. In Sec.~\ref{sec:22} a similar reduction leads to the derivation of the normal form associated to an essential codimension-two bifurcation of the system whose unfolding contains both of the previous scenarios. Section~\ref{sec:3} is then specifically devoted to the study of the anomalous regime, the formation of bright LSs and the different bifurcation structures associated with them. A similar study is presented in Sec.~\ref{sec:4}, this time focused on the normal regime, and on the formation and bifurcation structure of dark LSs. In Sec.~\ref{sec:5} we present some of the oscillatory and chaotic dynamics scenarios associated with both regimes. Finally, Sec.~\ref{sec:6} demonstrates the impact that the loss of spatial reversibility may have on the bifurcation structures previously discussed. The paper concludes with a brief summary in  Sec.~\ref{sec:7}.

\section{The stationary problem and spatial dynamics}\label{sec:1}
In this work, we focus on the bifurcation structure and stability of the steady states, and therefore on solutions of the stationary LL equation:
\begin{equation}\label{LLEsta}
i\nu \frac{d^2A}{dx^2} -(1+i\Delta)A+i|A|^2A+S=0,
\end{equation}
or, equivalently,
\begin{equation}\label{sta_real}
[\mathcal{L}+\mathcal{N}]\left[\begin{array}{c}
U \\ V
\end{array}\right]+\left[\begin{array}{c}
S \\ 0
\end{array}\right]=\left[\begin{array}{c}
0 \\ 0
\end{array}\right],
\end{equation}
where the linear ($\mathcal{L}$) and nonlinear ($\mathcal{N}$) operators are given by
\begin{align}
\mathcal{L}=\left[\begin{array}{cc}
-1&\Delta\\-\Delta &-1
\end{array}
\right]+\left[\begin{array}{cc}
0&-\nu\\ \nu &0
\end{array}\right]\partial^2_x,&& \mathcal{N}=(U^2+V^2)\left[\begin{array}{cc}
0&-1\\1&0
\end{array}
\right]\left[\begin{array}{c}
U\\V
\end{array}
\right].	
\end{align}
Here $A\equiv U+iV$. This stationary equation supports different types of steady states, such as spatially periodic, localized, and disordered states, as well as homogeneous or uniform states, which will be characterized in detail in the following section.

To fully understand the bifurcation structure of such states, it is essential to characterize their temporal linear stability. If $A_s$ is a stationary state of the system, i.e., a solution of Eq.~(\ref{LLEsta}), its temporal stability can be computed by solving the eigenvalue problem 
\begin{align}
	L\psi=&\sigma\psi, && L\equiv\mathcal{L}(A_s)+\mathcal{D}\mathcal{N}(A_s),
\end{align} 
obtained from the linearization of Eq.~(\ref{LLE}) about $A_s$, where $\mathcal{D}\mathcal{N}(A_s)$ is the functional derivative of $\mathcal{N}$ with respect to $A$, and $\sigma$ and $\psi$ are, respectively, the eigenvalues and eigenfunctions of $L$. Linear stability can only be determined analytically when an exact or approximate solution for $A_s$ is known, which is the case for the homogeneous state. When this is not possible, one can still determine stability, but this has to be done numerically by computing the eigenvalues of the Jacobian matrix obtained from $L$ after spatial discretization.  

\begin{figure}[!t]
\centering\includegraphics[scale=0.9]{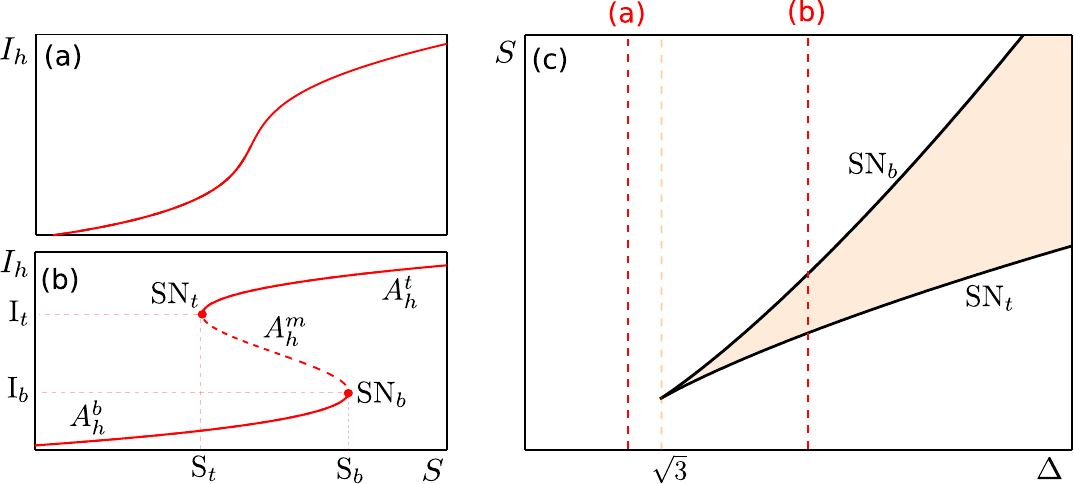}
\caption{Homogeneous steady state bifurcations and phase diagram. Depending on the value of $\Delta$, $I_h\equiv|A_h|^2$ is either (a) single-valued ($\Delta<\sqrt{3}$) or (b) triple-valued ($\Delta>\sqrt{3}$). These two bifurcation diagrams correspond to cuts through the phase diagram indicated by the vertical red dashed lines in panel (c), where we see how the folds SN$_{b,t}$ annihilate in a cusp bifurcation as $\Delta$ decreases to $\sqrt{3}$. Within the light orange region, the system has three homogeneous states.}
\label{cusp}
\end{figure}

\subsection{Homogeneous steady state and linear stability analysis}\label{sec:1.1}
The simplest steady state solution of Eq.~(\ref{LLEsta}) is obtained by imposing $\partial^2_xA=0$, and leads to the uniform or homogeneous steady state (HSS) solutions $A_h$, namely
\begin{align}\label{HSS0}
A_h=U_h+iV_h, && U_h=\frac{S}{1+(I_h-\Delta)^2}, && V_h\equiv \frac{(I_h-\Delta)S}{1+(I_h-\Delta)^2},
\end{align}
where $I_h\equiv|A_h|^2$ satisfies the classic cubic equation for optical bistability
\begin{equation}\label{HSS}
I_h^3-2\Delta I_h^2+(1+\Delta^2)I_h=S^2.
\end{equation}
For $\Delta<\sqrt{3}$, Eq.~(\ref{HSS}) is single-valued, and $I_h$ is a monotonic function of $S$ as shown in Fig.~\ref{cusp}(a). However, for $\Delta>\sqrt{3}$, $I_h$ is triple-valued [Fig.~\ref{cusp}(b)]. In the latter case, $I_h$ undergoes a pair of folds or saddle-node bifurcations SN$_{b,t}$ occurring at  
\begin{equation}
I_{t,b}\equiv|A_{t,b}|^2=\frac{2\Delta}{3}\pm\frac{1}{3}\sqrt{\Delta^2-3},
\end{equation}
with $I_b\le I_t$, which are created through a cusp or hysteresis bifurcation that takes place at $\Delta=\sqrt{3}$. These saddle-node bifurcations connect three branches of HSS solutions labeled $A_h^t$, $A_h^m$ and $A_h^b$. The cusp unfolding in the $(\Delta,S)-$phase diagram is shown in Fig.~\ref{cusp}(c). In the area between SN$_t$ and SN$_b$, the three HSS solutions $A_h^t$, $A_h^m$ and $A_h^b$ coexist. 

The temporal stability analysis of Eq.~(\ref{LLE}) around $A_h$ follows on considering small perturbations of the form $(U,V)=(U_h,V_h)+\epsilon (u_1,v_1)+\mathcal{O}(\epsilon^2)$, where $(u_1,v_1)=(a_q,b_q)e^{iqx+\sigma t}+c.c.$, and $|\epsilon|\ll1$. Inserting this ansatz in Eq.~(\ref{LLE}) and keeping terms of $\mathcal{O}(\epsilon)$, we obtain the perturbation growth rate
\begin{equation}\label{growth}
\sigma(q)=-1\pm\sqrt{4I_h\Delta-3I_h^2-\Delta^2+(4I_h-2\Delta)\nu q^2- q^4}. 
\end{equation}
Thus, $A_h$ is stable against perturbations of a given wavenumber $\bar{q}$ if Re$[\sigma(\bar{q})]<0$, and unstable otherwise. The instability threshold corresponds to Re$[\sigma(q)]_{q_c}=0$ and $d{\rm Re}[\sigma(q)]_{q_c}/dq=0$. These two conditions define the equations:
\begin{align}
q_c^4-\nu(4I_h-\Delta)q_c^2+3I_h^2+\Delta^2-4I_h\Delta+1=0,&& q_c(q_c^2-\nu(2I_h-\Delta))=0.	
\end{align}	
For homogeneous perturbations, or a system with no spatial extent (in practice a very small system), $q_c\equiv0$ and $A_h$ is always stable in the single-valued regime and exhibits bistability in the triple-valued regime (Fig.~\ref{cusp}). In larger systems, however, $q_c\neq0$ and the system undergoes a Turing instability \citep{turing_chemical_1952} at $I_h=I_T$. In an infinite system $q_c=\sqrt{\nu(2-\Delta)}$ and $I_T=1$. In the context of nonlinear optics this instability is often referred to as a modulational instability (MI). Note that the Turing instability exists whenever $\nu(2-\Delta)>0$, and therefore if $\nu=1$ ($\nu=-1$) the instability exists for $\Delta<2$ ($\Delta>2$).

\begin{figure}[!t]
	\centering\includegraphics[scale=0.9]{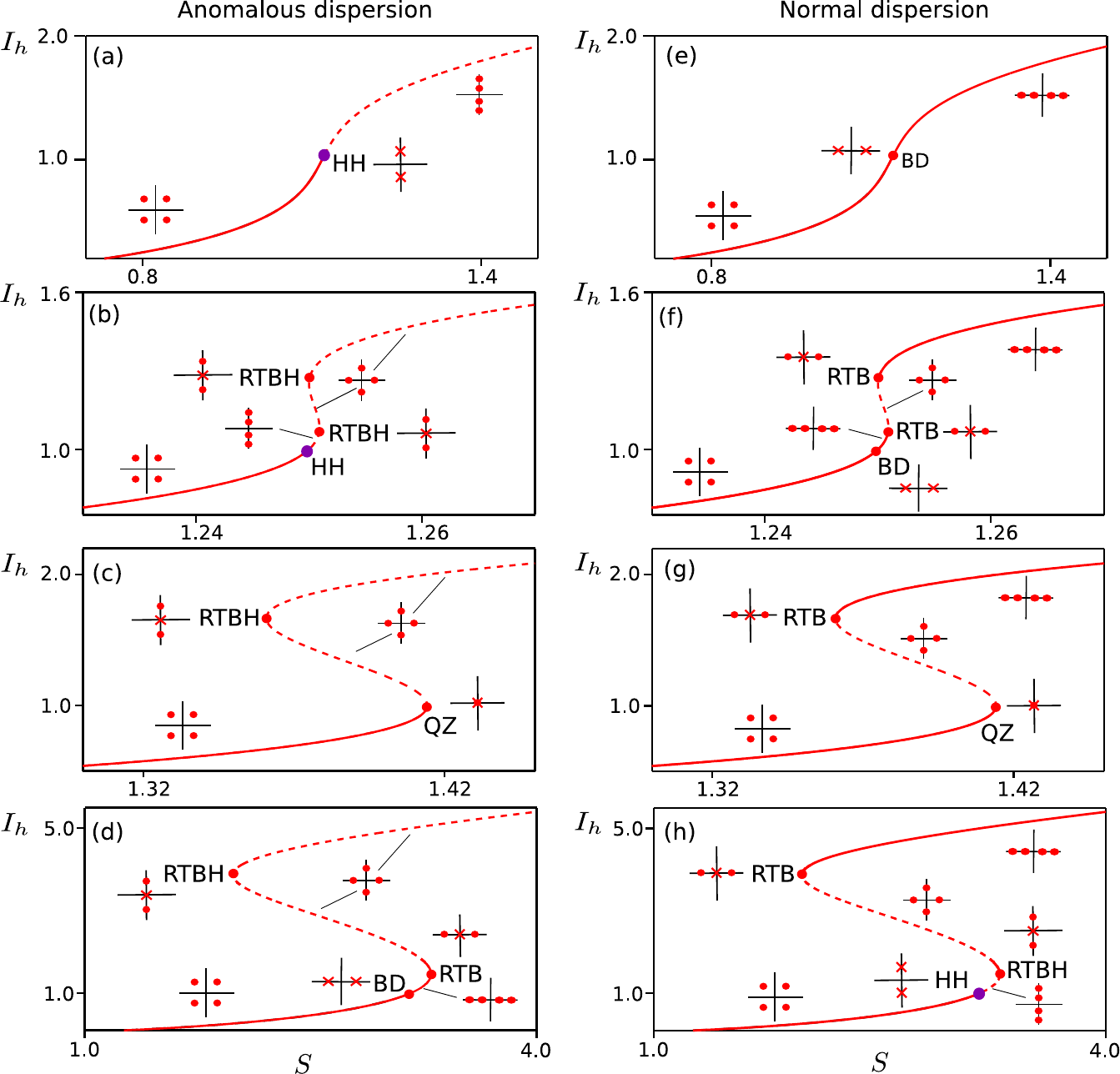}
	\caption{Homogeneous steady states and their spatiotemporal linear stability. Panels (a)-(d) show the bifurcation diagram of $A_h$ in the anomalous regime ($\nu=1$) for $\Delta=1.5,1.75,2.0,2.5$. Panels (e)-(h) show the same for the normal regime ($\nu=-1$). Solid and dashed lines represent stable and unstable states, respectively. The different pictograms show the corresponding spatial eigenvalue configurations from Fig.~\ref{unfolding_QZ}. The spatial bifurcation HH (see text) corresponds to the Turing instability and is marked with a purple dot. Adapted from \citet{parra-rivas_bifurcation_2018}.}
	\label{HSS_unfolding}
\end{figure}

In Fig.~\ref{HSS_unfolding}(a)-(d) we show how the stability of $A_h$ changes as $\Delta$ varies in the anomalous regime ($\nu=1$). The solid (dashed) lines represent stable (unstable) states, and the purple dot indicates the Turing instability. Similarly, panels (e)-(h) in Fig.~\ref{HSS_unfolding} show the stability of $A_h$ in the normal regime ($\nu=-1$) for the same values of $\Delta$. Note that in the anomalous regime $A_h^t$ is always unstable, but in the normal regime it is always stable. Thus, in the anomalous regime only $A_h^b$ can be (partially) stable, while in the normal dispersion regime $A_h^b$ and $A_h^t$ can both be stable in certain ranges of parameters, and the system is then said to exhibit bistability. This fact is essential for understanding the different types of LSs arising in each of these scenarios, as well as their bifurcation structure. 

\subsection{The spatial dynamics picture}\label{sec:1.2}

\begin{figure}[!t]
\centering\includegraphics[scale=0.9]{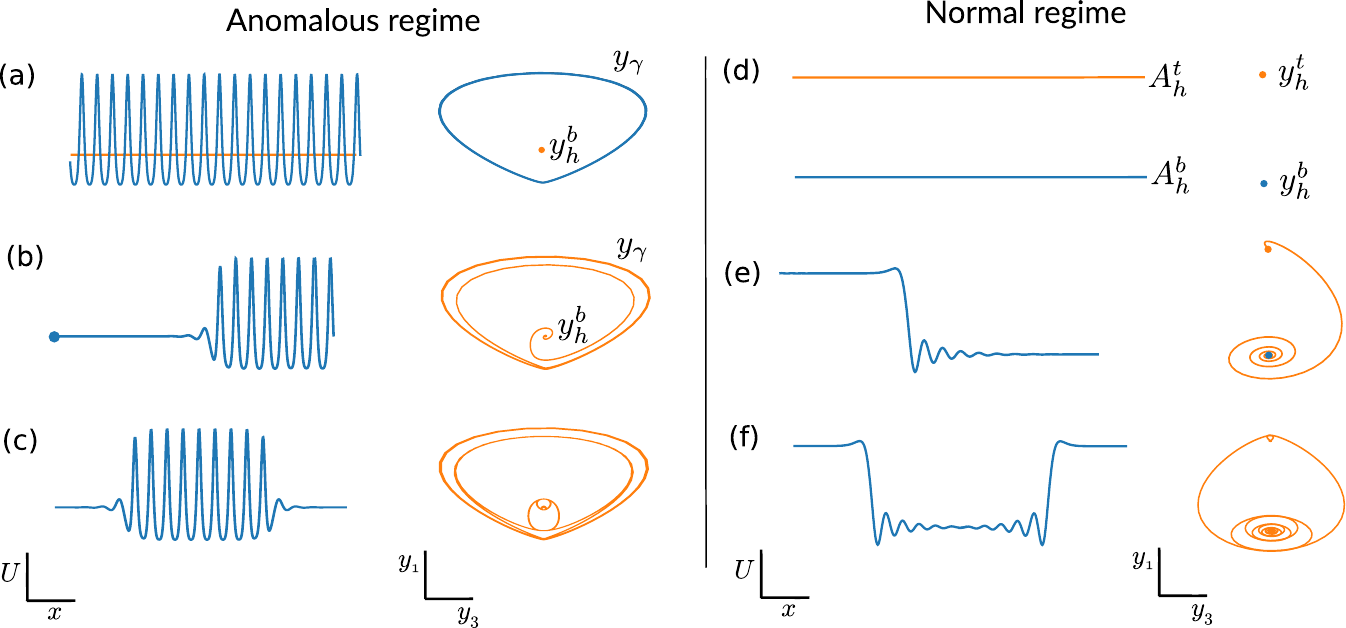}
\caption{Analogy between the stationary solutions of the Eq.~(\ref{LLE}) and the solution of the spatial dynamical system (\ref{SD}) in the anomalous and normal regime. For the solution of the LL equation we plot $U$ as a function of $x$, while in the spatial dynamics counterpart we consider its projection on the $(y_3,y_1)$ phase-plane.}
\label{fig_esquema}
\end{figure}

To understand the formation and origin of the different types of steady states arising in the system, it is convenient to recast the stationary equation (\ref{LLEsta}) as a four-dimensional dynamical system
\begin{align}\label{SD}
\frac{dy}{dx}=\mathcal{A}(\Delta)y+N(y;S), && y=(y_1,y_2,y_3,y_4)^T\equiv(U,V,U_x,V_x)^T, 
\end{align}
with
\begin{align}
\mathcal{A}(\Delta)\equiv\left[\begin{array}{cccc}0&0&1&0\\
0&0&0&1\\
\Delta&1&0&0\\
-1&\Delta&0&0\\
\end{array}\right], && N(y;S)\equiv\left[\begin{array}{c}0\\
0\\
-y_1y_2^2-y_1^3\\
-y_2y_1^2-y_2^3+S\\
\end{array}\right],
\end{align}
and then analyze its phase-space dynamics. In the context of pattern forming systems, this technique is usually known as the {\it spatial dynamics} approach, and it allows one to understand the emergence of LSs from a dynamical systems perspective \citep{haragus_local_2011}. Equation~(\ref{LLE}) is invariant under the spatial reflection $x\rightarrow -x$, which leads to the invariance of the dynamical system (\ref{SD}) under the involution $$R (x,y_1,y_2,y_3,y_4)\mapsto(-x,y_1,y_2,-y_3,-y_4).$$ When this symmetry holds, the system is said to be {\it spatially reversible}. The equivalence between the spatial and temporal formulations permits one to establish a correspondence between the solutions of Eq.~(\ref{LLEsta}) and those of the dynamical system (\ref{SD}). This duality is shown schematically in Fig.~\ref{fig_esquema} for the anomalous and normal regimes. For each regime, the left column shows the typical steady state solutions of Eq.~(\ref{LLE}), whereas the right column shows the equivalent orbits in the $(y_3,y_1)$-phase plane projection associated with Eq.~(\ref{SD}). In this picture, the homogeneous solution $A_h^b$ corresponds to a fixed point $y_h^b=(U_h^b,V_h^b,0,0)$, while a spatially periodic state corresponds to a limit cycle $y_\gamma$. An interesting situation arises when different types of states (e.g., a fixed point and a cycle) coexist for the same set of parameters. Different types of {\it heteroclinic orbits} can then arise, corresponding to DWs or front solutions of Eq.~(\ref{LLEsta}), leading to the complex scenarios to be explained below.

In the anomalous regime, bistability between the subcritical periodic Turing state $P$ and $A_h^b$ [see Fig.~\ref{fig_esquema}(a)] leads to the emergence of fronts like that shown in Fig.~\ref{fig_esquema}(b), corresponding to a heteroclinic orbit connecting $y_h^b$ and $y_\gamma$. These connections form as a result of a transverse or robust intersection between the unstable manifold of $y_h^b$ ($W^u[y_h^b]$) and the stable manifold of $y_\gamma$ ($W^s[y_\gamma]$); the robustness of this intersection is in turn a consequence of the dimensions of these manifolds, as further explained in \citet{knobloch_spatial_2015}. Furthermore, spatial reversibility implies a similar intersection between $W^u[y_\gamma]$ and $W^s[y_h^b]$, and hence the presence of a heteroclinic {\it cycle}; homoclinic orbits in $W^s[y_h^b]\cap W^u[y_h^b]$ accumulate on this cycle. An example of such an orbit is shown in Fig.~\ref{fig_esquema}(c), where the trajectory rotates several times around $y_\gamma$ before returning to $y_h^b$. Solutions of this type correspond to {\it localized patterns} (LPs) containing a long plateau where the solution resembles the spatially periodic pattern shown in Fig.~\ref{fig_esquema}(b). Each rotation around $y_\gamma$ generates an additional peak in the profile of the LP. These orbits approach or leave $A_h^b$ in an oscillatory manner, leading to the appearance of oscillatory tails in the LP profile, and correspond to {\it Shilnikov} or {\it wild} homoclinic orbits \citep{champneys_when_2007,homburg_homoclinic_2010}. In contrast, orbits where the behavior around the fixed point is monotonic are known as {\it tame} homoclinic orbits, and correspond to a {\it spike} \citep{verschueren_model_2017}.

The normal dispersion regime is very different as $A_h^b$ and $A_h^t$ can coexist in a stable way [see Fig.~\ref{fig_esquema}(d)]. As a result, heteroclinic orbits can arise from the intersection between $W^s[y_h^b]$ and $W^u[y_h^t]$, forming the DW shown in Fig.~\ref{fig_esquema}(e).  As in the anomalous regime, spatial reversibility is responsible for the formation of a variety of homoclinic orbits such as that shown in Fig.~\ref{fig_esquema}(f). The formation of such LSs can be physically understood in terms of DWs that lock to one another, a mechanism which will be presented in Sec.~\ref{sec:4}.  

\begin{figure}[!t]
	\centering\includegraphics[scale=0.9]{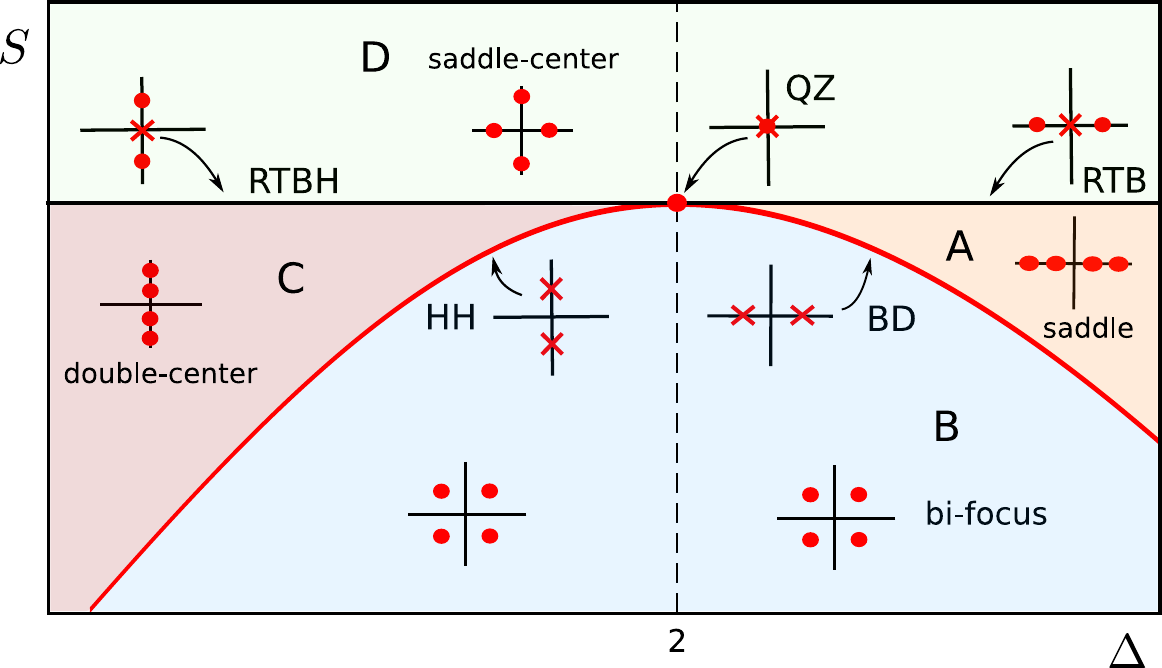}
	\caption{Schematic unfolding of the QZ point in the $(\Delta,S)$-parameter space. Below $\Delta=2$, SN$_b$ corresponds to a RTBH bifurcation, and the Turing instability to HH. At the QZ point ($\Delta=2$), these bifurcations collide, and for $\Delta>2$ SN$_b$ becomes a RTB bifurcation and HH turns into a BD transition. These four lines organize the different type of equilibria of the system. Adapted from \citet{champneys_homoclinic_1998}.}
	\label{unfolding_QZ}
\end{figure}

The origin of the previous trajectories and their behavior near the fixed point $y_h$ (i.e., $A_h$) can be understood by analyzing the spectrum of the linear operator $\mathcal{A}+\mathcal{D}N$ evaluated at $y_h$, which consists of the four spatial eigenvalues  
\begin{equation}\label{eq_spa_eigen}
\lambda=\pm\sqrt{(\Delta-2I_h)\nu\pm\sqrt{I_h^2-1}}.
\end{equation}
These eigenvalues lead to four different equilibrium configurations (regions A-D) depending on the values of the parameters $\Delta$ and $I_h$ (or $S$). These four configurations are depicted in the phase diagram shown in Fig.~\ref{unfolding_QZ}, and are defined as follows: in A, $y_h$ is a saddle ($s$) with eigenvalues $\lambda_{1,2}=\pm a_1$, $\lambda_{3,4}=\pm a_2$; in B, $y_h$ is a bi-focus (bi-$f$) with the quartet of complex eigenvalues $\lambda_{1,2,3,4}=\pm a_0 \pm ib_0$; in C $y_h$ is a double-center ($dc$) with imaginary eigenvalues $\lambda_{1,2}=\pm ib_1$, $\lambda_{3,4}=\pm ib_2$; and in D, $y_h$ is a saddle-center ($sc$) with two real and two purely imaginary eigenvalues $\lambda_{1,2}=\pm a_0$, $\lambda_{3,4}=\pm ib_0$.

The transition from one region to an adjacent one occurs via the following codimension-one bifurcations or transitions:
\begin{itemize}
	\item A Belyakov-Devaney (BD) transition occurs between
	regions A and region B. At this point the spatial eigenvalues are real: $\lambda_{1,2}=\pm a$, $\lambda_{3,4}=\pm a$.
	\item The transition between region B and region C is via a Hamiltonian-Hopf (HH) bifurcation,
	with purely imaginary eigenvalues: $\lambda_{1,2}=\pm iq_c$, $\lambda_{3,4}=\pm iq_c$.
	\item The transition between region A and region D is via a reversible Takens-Bogdanov (RTB) bifurcation with 
	eigenvalues $\lambda_{1,2}=\pm a$, $\lambda_{3}=\lambda_{4}=0$. 
	\item The transition between region C and region D is via a reversible 
	Takens-Bogdanov-Hopf (RTBH) bifurcation with eigenvalues $\lambda_{1,2}=\pm ib$, $\lambda_{3}=\lambda_{4}=0$.
\end{itemize}
Note that the spatial eigenvalues given by Eq.~(\ref{eq_spa_eigen}) can also be obtained from Eq.~(\ref{growth}) by imposing $\sigma(-i\lambda)=0$. As a result, the HH bifurcation corresponds to a Turing instability, while RTB and RTBH correspond to SN$_{b,t}$. This scenario is generic for reversible four-dimensional dynamical systems \citep{devaney_reversible_1976,champneys_homoclinic_1998,haragus_local_2011}, and it is organized by a quadruple zero (QZ) codimension-two bifurcation satisfying $\lambda_1=\lambda_2=\lambda_{3}=\lambda_{4}=0$ \citep{iooss_codimension_1995}. Here, the QZ occurs at $(\Delta,S)=(2,\sqrt{2})$, and it organizes the appearance of the different types of steady state solutions in the anomalous and normal regimes \citep{parra-rivas_dynamics_2014,godey_stability_2014}. The transition between these different scenarios is shown in Fig.~\ref{HSS_unfolding}(a)-(d) for the anomalous regime, where $I_h$ is plotted as a function of $S$ for different representative values of $\Delta$. Figure~\ref{HSS_unfolding}(e)-(h) shows the transition between the different regimes for the  normal regime and the same values of $\Delta$.

\section{Weakly nonlinear localized state solutions}\label{sec:2}

Close to the different spatial bifurcations discussed in Sec.~\ref{sec:1}, one can compute weakly nonlinear states using different approaches. One method consists in deriving the normal form associated to the dynamical system (\ref{SD}) around each of the spatial bifurcations, and solve the truncated system \citep{godey_bifurcation_2017}. However, one can also follow another approach where such weakly nonlinear states are obtained by applying multiscale perturbation theory to solve Eq.~(\ref{sta_real}). In this section, we review the main results that one obtains using this last method, and we refer to \citet{burke_classification_2008,parra-rivas_dark_2016,parra-rivas_bifurcation_2018} for a more detailed description.
 
Our two main bifurcation points of interest are the Turing bifurcation point (i.e., an HH spatial bifurcation) and the saddle-node bifurcation points SN$_{b,t}$. In the neighborhood of such bifurcations, weakly nonlinear time-independent states are captured by the ansatz:
$$A(x)-A_h\sim \epsilon Z(X)e^{iq_cx}+c.c.,$$
where $\epsilon\ll1$ measures the parameter distance from the bifurcation, $q_c$ is the characteristic wavenumber of the marginal mode at the bifurcation ($q_c=0$ for the fold, and $q_c\neq0$ for HH) and $Z$ is an envelope function describing spatial modulation occurring at a larger scale $X=\epsilon^\alpha x$, where $\alpha$ depends on each specific case. In the following, we split the stationary solutions as $\left(U,V\right)^T=\left(U_h,V_h\right)^T+\left(u(x,X),v(x,X)\right)^T$, to separate the homogeneous part of the problem from the space-dependent one. 

\subsection{Weakly nonlinear states near the Hamiltonian-Hopf bifurcation}\label{sec:2.1}
To compute the weakly nonlinear states near HH, we fix $\Delta$, consider $S=S_c+\delta \epsilon^2$, and propose the expansions 
$\left(U_h,V_h\right)^T=\left(U_c,V_c\right)^T+\epsilon^2 \left(U^h_2,V^h_2\right)^T+\cdots$, and  
$ \left(u,v\right)^T=\epsilon \left(u_1,v_1\right)^T+\epsilon^2 \left(u_2,v_2\right)^T +\epsilon^3 \left(u_3,v_3\right)^T+\cdots$,
where $(u_i,v_i)^T$ depend on both the short scale $x$ and the long scale $X\equiv\epsilon x$. Inserting these expansions into Eq.~(\ref{sta_real}), keeping the terms of the same order in $\epsilon$, and solving the resulting linear equations, we conclude that the asymptotic solution we are looking for can be written as 
\begin{align}\label{wild_asymp}
(U,V)\sim(U_c,V_c)^T+\frac{S-S_c}{\delta}(U_2,V_2)+\sqrt{\frac{S-S_c}{\delta}}(u_1,v_1),&& S-S_c\rightarrow 0,
\end{align}
where $(U_c,V_c)^T$ is given by Eq.~(\ref{HSS0}) evaluated at $I_h=I_c$. Here $(U_2,V_2)^T$ represents the leading order correction to the homogeneous solution, namely 
\begin{equation}
(U_2,V_2)^T=\frac{\delta}{{\left(\Delta^{2} - 2 \, \Delta +2\right)} {\left(\Delta-2\right)}}\left(\Delta^{2},2-\Delta^{2} - \Delta\right)^T,
\end{equation}
while the space-dependent correction reads
\begin{equation}
(u_1,v_1)^T=2\left(\frac{\Delta}{2-\Delta},1\right)^T Z(X)\,{\rm cos}(q_c x+\varphi).
\end{equation}
The amplitude $Z(X)$ is the solution of the time-independent normal form around HH,
\begin{equation}\label{amplitude_HH}
C_1 Z_{XX}+\delta C_2Z+C_3Z^3=0,
\end{equation}
with the coefficients
\begin{align}
C_1=-\frac{2 \, {\left(\Delta^{2} - 2 \, \Delta + 2\right)}}{\Delta - 2},&&C_2=
\frac{2 \, {\left(\Delta^{2} - 2 \, \Delta +
		2\right)}^{\frac{3}{2}}}{{\left(\Delta - 2\right)}^{4}},&& C_3=\frac{4 \, {\left(\Delta^{2} - 2 \, \Delta +
		2\right)}^{2} {\left(30 \, \Delta - 41\right)}}{9 \, {\left(\Delta -
		2\right)}^{6}}.
\end{align}
When $Z(X)\equiv Z$, Eq.~(\ref{amplitude_HH}) leads to the constant solution $Z=\sqrt{-\delta C_2/ C_3}$, which corresponds to the spatially periodic pattern state
\begin{equation}\label{Turing_pattern}
(U,V)^T-(U_h,V_h)^T\sim 2\left(\frac{\Delta}{2-\Delta},1\right)^T \sqrt{\displaystyle\frac{C_2}{C_3}(S_c-S)}\,\textnormal{cos}\left(q_cx+\varphi\right),
\end{equation}
where $\varphi$ is an arbitrary phase. Note that these solutions exist whenever $\Delta<2$, and arise from HH sub- or supercritically depending on the sign of $C_3$. The case with $C_3>0$ corresponds to $\Delta>41/30$, and leads to a subcritical emergence of the pattern from HH (i.e., the  pattern bifurcates towards $S<S_c$). In contrast, for $\Delta<41/30$ the pattern arises supercritically, i.e., towards $S>S_c$. These results agree with those obtained previously by different authors when studying the dynamics of periodic Turing patterns near the HH point \citep{lugiato_spatial_1987,miyaji_bifurcation_2010,perinet_eckhaus_2017,godey_bifurcation_2017}.\\

In the subcritical regime, solutions with a large scale modulation $Z\equiv Z(X)$ are present, and these are given by 
\begin{equation}
Z(X)=\sqrt{\displaystyle\frac{-2\delta C_2}{C_3}}\,\textnormal{sech}\left(\displaystyle\sqrt{-\nu\delta C_2/C_1}X\right),
\end{equation}
corresponding to 
\begin{equation}\label{WNL_Wild_LP}
(U,V)^T-(U_h,V_h)^T\sim 2\left(\frac{\Delta}{2-\Delta},1\right)^T
\sqrt{\displaystyle\frac{-2 C_2(S-S_c)}{C_3}}\,\textnormal{sech}\left(\displaystyle\sqrt{\frac{-\nu C_2(S-S_c)}{C_1}}x\right)\textnormal{cos}\left(q_cx+\varphi\right).
\end{equation}
The spatial phase $\varphi$ of the background periodic pattern remains arbitrary, and there is no locking with the envelope at any finite order in $\epsilon$. However, calculations beyond all orders predict that two specific values of $\varphi=0,\pi$ are selected, both preserving the reversibility symmetry $(x,A)\rightarrow(-x,A)$ of Eq.~(\ref{LLE}) \citep{melbourne_derivation_1998,burkeknobloch2006,kozyreff_asymptotics_2006,chapman_exponential_2009,kozyreff_localized_2012}. Thus there are two types of localized weakly nonlinear solutions, one with a maximum at the center of the domain ($x=0$), corresponding to $\varphi=0$, and another with a minimum at $x=0$, associated with $\varphi=\pi$. In the following we label such families of solutions as $\Gamma_0$, and $\Gamma_\pi$, respectively.

\subsection{Weakly nonlinear states near the saddle-node bifurcations SN$_{b,t}$}\label{sec:2.2}
Next, we look for weakly nonlinear solutions around the saddle-node bifurcations ${\rm SN}_r\equiv{\rm SN}_{b,t}$, and focus on the case where they correspond to a RTB bifurcation. To do so, we again propose $S\approx S_{r}+\delta\epsilon^2$, with $S_r\equiv S_{b,t}$, and the asymptotic expansions $\left(U_h,V_h\right)^T=\left(U_{r},V_{r}\right)^T+\epsilon \left(U^h_1,V^h_1\right)^T+\epsilon^2 \left(U^h_2,V^h_2\right)^T+\cdots$, and  
$ \left(u,v\right)^T=\epsilon \left(u_1,v_1\right)^T+\epsilon^2 \left(u_2,v_2\right)^T +\cdots$,
where this time $(u_i,v_i)^T$ depend only on the long scale $X\equiv \sqrt{\epsilon}x$. Proceeding similarly as in the previous case, we can compute asymptotic weakly nonlinear LSs, which take the leading order form
\begin{equation}\label{tame_asymp}
(U,V)^T
= (U_{r},V_{r})^T+\sqrt{\frac{S-S_{r}}{\delta}}
(U^h_{1}+u_1,V^h_{1}+v_1)^T,
\end{equation}
where $(U_{r},V_{r})^T$ is given by Eq.~(\ref{HSS}) evaluated at $I_h=I_r\equiv I_{b,t}$,
and $(U^h_1,V^h_1)$ is the leading order correction to $A_h$, namely
\begin{align}
(U^h_1,V^h_1)^T=\sqrt{\delta}\mu_r(1,\eta_r)^T, &&\eta_r= -\displaystyle\frac{1}{2}(\Delta-I_{r}-2U_{r}^2),&&\mu_r=\mu^s_r\sqrt{\frac{\Upsilon_r^s}{|\Upsilon_r|}}, 
\end{align}
with $\Upsilon_r=3\eta_r^2V_{r}+2\eta_r U_{r}+V_{r}$,  $\Upsilon_r^s={\rm sign}(\Upsilon_r)$, $\mu^s_b=-1$, and $\mu^s_t=1$.

The space-dependent contribution is given by
$(u_1,v_1)^T=(U_1^h,V^h_1)Z(X)$,
where the amplitude $Z(X)$ is a solution of the time-independent normal form:
\begin{equation}\label{psi.eq.1_down}
\displaystyle\frac{\nu\eta_r\mu_r}{\sqrt{\delta}}Z_{XX}+2Z+Z^2=0.
\end{equation}
This equation supports solutions of the form 
\begin{equation}
Z(X)=-3\,\textnormal{sech}^2\left(\displaystyle\frac{1}{2}\sqrt{\frac{-2\sqrt{\delta}}{\nu\eta_r\mu_r^s}\sqrt{\frac{|\Upsilon_r|}{\Upsilon_r^s}}}X\right),
\end{equation}
corresponding to tame weakly nonlinear LSs
\begin{equation}\label{tame_LS}
(U,V)^T-(U_h,V_h)^T\sim -3\mu_r\sqrt{S-S_r}(1,\eta_r)^T{\rm sech}^2\left(\frac{1}{2}\sqrt{C_r\sqrt{|S-S_r|}}x \right),
\end{equation}
where the HSS term $(U_h,V_h)^T$ contains the contribution of $(U_{b,t},V_{b,t})^T$ and $(U_{1}^h,V_{1}^h)^T$, 
and 
\begin{align}
C_r\equiv\frac{-2}{\nu\eta_r\mu_r^s}\sqrt{\frac{|\Upsilon_r|{\rm sign}(S-S_r)}{\Upsilon_r^s}}
\end{align}
is a positive coefficient.
Close to SN$_b$ (i.e., $r=b$), and SN$_t$ (i.e., $r=t$) $C_r$ reduces to $C_b=2\sqrt{|\Upsilon_b|}/\nu\eta_b$, and $C_t=-2\sqrt{|\Upsilon_t|}/\nu\eta_t$, respectively. Thus, tame homoclinic orbits of the form (\ref{tame_LS}) arise from the spatial RTB bifurcation SN$_b$ in the anomalous regime $(\nu=1)$ whenever $\Delta>2$, and in the normal regime $(\nu=-1)$ whenever $\sqrt{3}<\Delta<2$. In contrast, such states emerge from SN$_t$ only in the normal regime, but for any value of $\Delta>\sqrt{3}$.

When the folds SN$_{b,t}$ correspond to RTBH bifurcations the situation is rather more delicate \citep{haragus_local_2011}. In this case, new states, commonly known as {\it generalized solitary waves}, may be present. These states are biasymptotic to a spatially periodic state of constant but arbitrarily small amplitude. Embedded among these generalized solitary states are true homoclinic states or exponentially localized states with no oscillations in their tail, as described by \citet{kolossovski_multi-pulse_2002}. A proper computation of these states requires the application of a careful normal form approach to Eq.~(\ref{SD}), as done by \citet{godey_bifurcation_2017} in the context of Eq.~(\ref{LLE}). However, as found by \citet{gandhi_spatially_2018}, the weakly nonlinear solution (\ref{tame_LS}), obtained through formal multiscale perturbation analysis that ignores the center eigenvalues, may provide a good approximation to such states provided one replaces $C_r$ in Eq.~(\ref{tame_LS}) by $|C_r|$. As far as we know, these types of states have not been studied in detail in the present context, and are left for a future work.

\section{The origin of all localized structures: the quadruple-zero point}\label{sec:22}

To completely understand the origin of the LSs, the different bifurcation scenarios, and the transition between them in both the anomalous and normal regimes, it is essential to unveil the dynamics emerging nearby the QZ codimension-two bifurcation.  The first systematic study of the dynamical features of this point was carried out by \citet{iooss_codimension_1995} in a scenario involving a trivial state. Here, however the QZ does not take place on a trivial state but on a non-trivial one. In this section, we reduce the LL equation (\ref{LLE}) to the unfolded normal form associated with the QZ bifurcation. We show that this last equation captures the main local features of the system about QZ. In the following we focus on the anomalous regime and therefore fix $\nu=1$.

\begin{figure}[!t]
	\centering\includegraphics[scale=0.9]{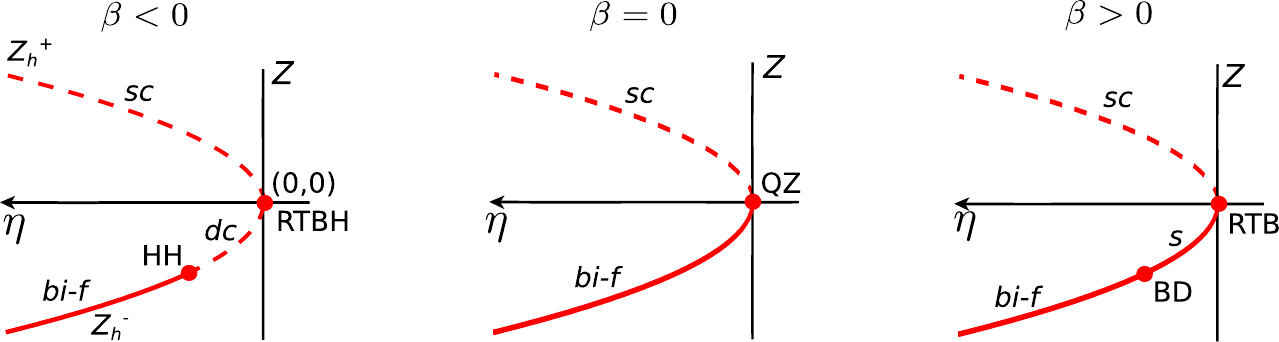}
	\caption{Homogeneous steady state solutions of Eq.~(\ref{normalQZ}), and possible unfolding of the QZ bifurcation as a function of $\beta$. The red line shows the two HSS solution branches $Z_h^{\pm}$ separated by a fold at $\eta=0$. For $\beta<0$, $Z_h$ undergoes a HH bifurcation. At $\beta=0$ HH collides with the fold leading to the QZ point. For $\beta>0$, HH becomes a BD and RTBH becomes RTB. Solid (dashed) lines correspond to stable (unstable) states. }
	\label{diaQZ}
\end{figure}

The HSS $A_h$ undergoes a QZ bifurcation at $(U_h^{Q},V_h^Q)^T=(1/\sqrt{2},-1/\sqrt{2})$ at the parameter space point $(\Delta,S)=(\Delta_Q,S_Q)\equiv(2,\sqrt{2})$. To explore the dynamics of the system around QZ we introduce a small parameter $\epsilon$ measuring the distance from this point, $\Delta=\Delta_Q+\epsilon^2\beta$, and write $S=S_Q+\epsilon^4\eta$ and $(U(x,t),V(x,t))^T=(U_h^Q,V_h^Q)^T+(u(x,t),v(x,t))$. We also introduce the slow scales $X\equiv\epsilon x$ and $T\equiv\epsilon^4 t$, and the scaling $(u,v)\rightarrow\epsilon^4(u,v)$. 
Finally we expand the deviation from QZ as the power series in $\epsilon$
\begin{equation}
(u,v)=\epsilon (u_1,v_1)+\epsilon^2 (u_2,v_2)+\epsilon^3(u_3,v_3)+\epsilon^4 (u_4,v_4)+\cdots,
\end{equation}
with  
$(u_i,v_i)$ depending on $X$ and $T$.
Inserting this expansion in Eq.~(\ref{LLE}) and keeping all the terms at the same order in $\epsilon$, one obtains at $\mathcal{O}(\epsilon^4)$ the required normal form about the QZ point,
\begin{equation}\label{normalQZ}
Z_T=-Z_{XXXX}+\beta Z_{XX}+Z^2+\eta,
\end{equation}
where $\beta\propto \Delta-\Delta_Q$, $\eta\propto S-S_Q$ are the two unfolding parameters, and $(u_4,v_4)^T=(\xi_1,\xi_2)^TZ(X,T)$, with $\xi_i\in\mathbb{R}$.
This equation has gradient structure, and therefore temporal dynamical states are excluded. The simplest steady state is the HSS $Z_h$ given by $Z_h=\pm\sqrt{-\eta}$, composed of two solution branches $Z_h^{\pm}$ connected by a fold at $\eta=0$ (see Fig.~\ref{diaQZ}). For $\beta<0$, the HSS undergo a Turing bifurcation at $Z_c=-\beta^2/8$ that gives rise to spatially periodic states with wavenumber $q_c^2=-\beta/2$, while the fold corresponds to a saddle-node bifurcation. The stability of $Z_h$ against spatiotemporal perturbations is shown in Fig.~\ref{diaQZ} using solid (dashed) lines for stable (unstable) states.

The time-independent version of Eq.~(\ref{normalQZ}) can be recast into the dynamical system
\begin{align}\label{SD2}
\frac{dz}{dX}=\mathcal{A}(\beta)z+N(z;\eta), && z=(z_1,z_2,z_3,z_4)^T\equiv(Z,Z_X,Z_{XX},Z_{XXX})^T, 
\end{align}
with
\begin{align}
\mathcal{A}(\beta)\equiv\left[\begin{array}{cccc}
0&1&0&0\\
0&0&1&0\\
0&0&0&1\\
0&0&\beta&0\\
\end{array}\right], && N(z;\eta)\equiv\left[\begin{array}{c}0\\
0\\
0\\
z_1^2+\eta\\
\end{array}\right].
\end{align}

The linearization of this dynamical system about $Z_h$ leads to a spatial eigenspectrum consisting of the four eigenvalues satisfying $\lambda^4-\beta\lambda^2-2Z_h=0$, i.e.,
\begin{equation}
\lambda=\pm\sqrt{\frac{\beta\pm\sqrt{\beta^2+8Z_h}}{2}}.
\end{equation}
Depending on the value of $\beta$, three possible scenarios may occur which are schematically described in Fig.~\ref{diaQZ}. For $\beta<0$, $Z^-_h$ encounters a HH at $(\eta_c,Z_c)=-(\beta^4/64,\beta^2/8)$, such that $Z^-_h$ is a bi-focus for $\eta<\eta_c$ and double center for $\eta>\eta_c$. The fold encountered at $\eta=0$ corresponds to a RTBH bifurcation with eigenvalues $\lambda_{1,2,3,4}=(\pm i\sqrt{|\beta|},0,0)$ from where the saddle-center $Z^+$ arises. In this context, spatially periodic solutions may bifurcate from HH subcritically together with the two families of wild homoclinic orbits corresponding to $\Gamma_{0,\pi}$ as described in Sec.~\ref{sec:2.1}. For $\beta=0$, HSS encounters the QZ bifurcation at $(\eta,Z_h)=(0,0)$ as shown in Fig.~\ref{diaQZ}. For $\beta>0$ HH has become a BD, and the fold a RTB bifurcation with eigenvalues $\lambda_{1,2,3,4}=(\pm\sqrt{|\beta|},0,0)$. From this last point tame homoclinic orbits may arise as described in Sec.~\ref{sec:2.2}. Therefore, the normal form (\ref{normalQZ}) captures the main spatial dynamical features of the LL equation around the QZ bifurcation that takes place at the fold SN$_b$, as depicted in Figs.~\ref{HSS_unfolding}(b)-(d).
 
Note that the change of variable $Z\rightarrow Z_h+Z$ transforms Eq.~(\ref{normalQZ}) into the quadratic SH equation studied by \citet{buffoni_bifurcation_1996}. Consequently most of the results found in that work apply here as well, although they require reinterpretation. A complete understanding of Eq.~(\ref{normalQZ}) thus requires further analysis that is beyond the scope of the present paper.

\section{Localized structures in the anomalous regime: Homoclinic and foliated snaking}\label{sec:3}
\begin{figure}[!t]
\centering\includegraphics[scale=0.8]{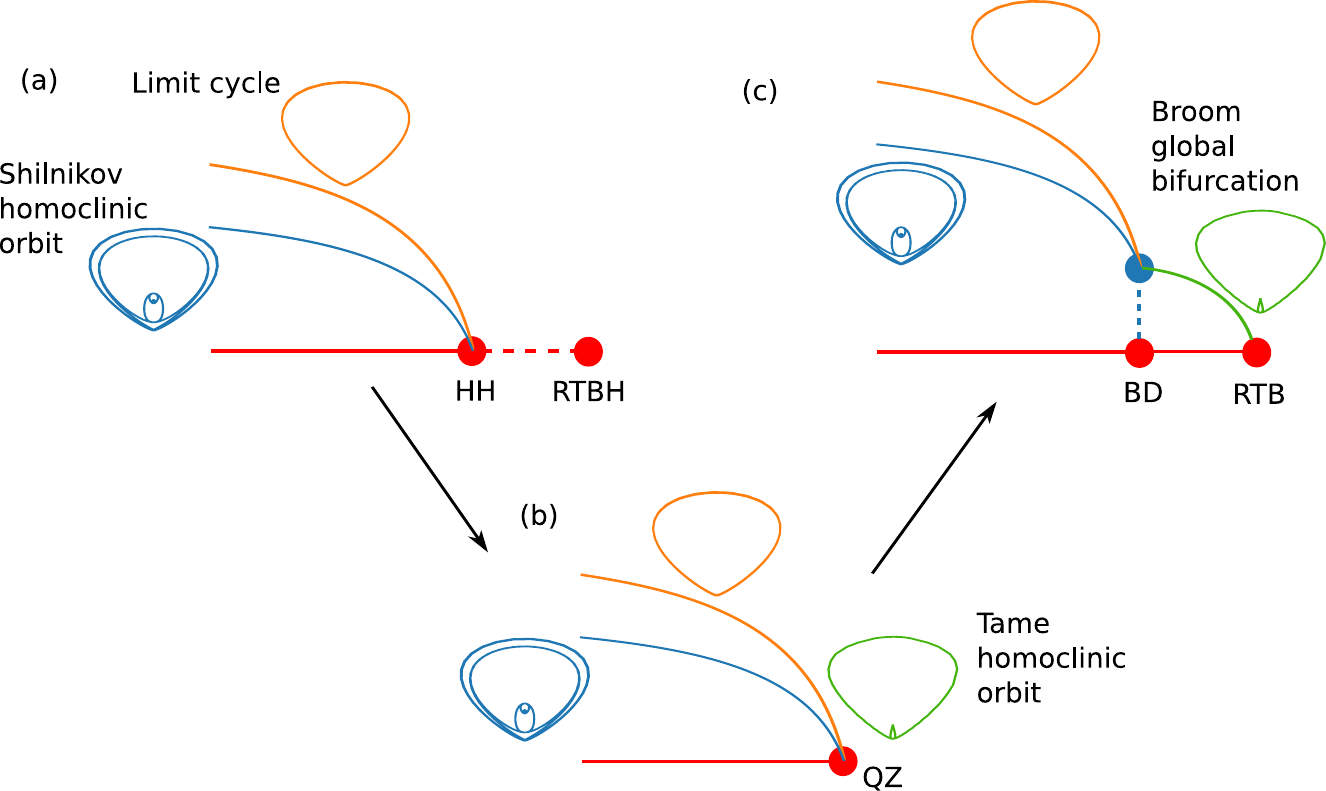}
\caption{Schematic representation of the different bifurcation scenarios around QZ in the anomalous regime that may arise as $\Delta$ varies. In (a) $\Delta<2$, and a spatially periodic state (i.e., a limit cycle) arises from HH together with two families of LPs (i.e., Shilnikov homoclinic orbits). In (b) $\Delta=2$ and a QZ occurs when HH collides with the saddle-node RTBH bifurcation. At the QZ point a global homoclinic bifurcation takes place, where the periodic pattern becomes a spike. In (c) $\Delta>2$ and the LPs and periodic patterns undergo a global bifurcation at BD, leaving a single spike, which then survives until the saddle-node RTB where it disappears. }
\label{unfoldingQZ}
\end{figure}

In this section we discuss the origin and bifurcation structure of the LSs arising in the anomalous dispersion regime ($\nu=1$). The different bifurcation scenarios in this regime are organized by the QZ point, as shown schematically in Fig.~\ref{unfoldingQZ}, and  discussed in more detail in what follows. In Fig.~\ref{unfoldingQZ}(a), when $41/30<\Delta<2$, a spatially periodic pattern arises subcritically from HH together with two families of LPs ($\Gamma_{0,\pi}$) that are ultimately responsible for the snakes-and-ladders bifurcation structure of the LPs in the snaking regime. This snakes-and-ladders structure is also linked to the bifurcation features of the periodic pattern \citep{parra-rivas_bifurcation_2018}. This scenario is presented in Secs.~\ref{sec:3.1} and \ref{sec:3.2}. When $\Delta=2$, the HH collides with the RTBH (i.e., SN$_b$), leading to a QZ, see Fig.~\ref{unfoldingQZ}(b). At this point a global homoclinic bifurcation takes place, and the spatially periodic and localized patterns come together forming spike LSs (i.e., a tame homoclinic orbit) for $\Delta>2$ \citep{parra-rivas_bifurcation_2018}. Figure~\ref{unfoldingQZ}(c) shows how the spike LSs bifurcate from RTB and persist until the BD point, where they are destroyed in a {\it broom} homoclinic bifurcation, leading to the emergence of spatially periodic pattens and LPs. These two scenarios are discussed in Sec.~\ref{sec:3.3}.

\subsection{Bifurcation structure of the periodic Turing patterns} \label{sec:3.1}
\begin{figure}[!t]
	\centering\includegraphics[scale=0.8]{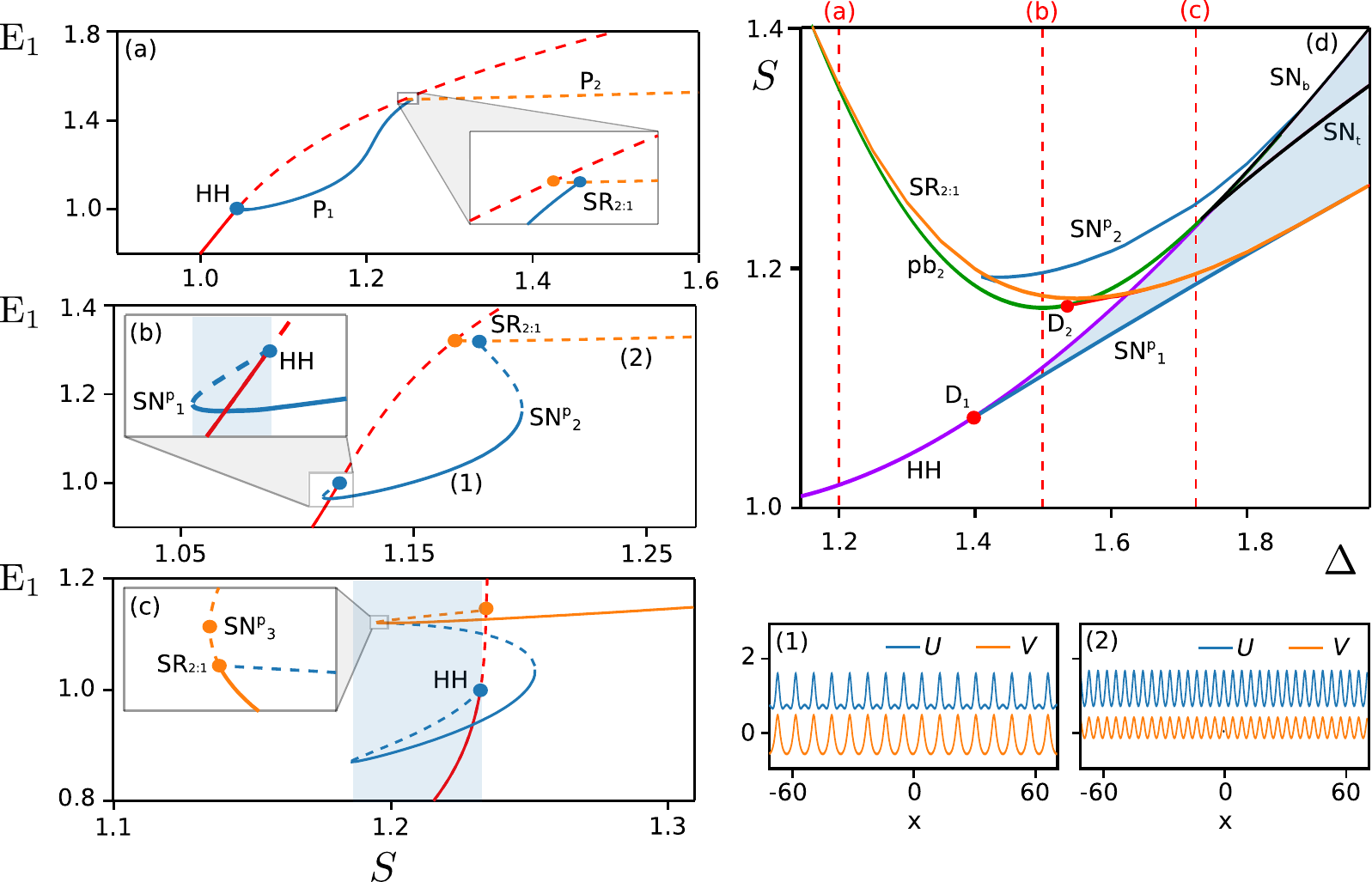}
	\caption{Bifurcation diagrams associated with the primary Turing pattern (P$_1$, $q=q_c$) emerging from HH for three representative values of $\Delta$, namely $\Delta=1.2$ in (a), $\Delta=1.5$ in (b) and $\Delta=1.72$ in (c). Panel (d) shows the phase diagram in the $(\Delta,S)$-parameter space, where the main bifurcation curves relevant to spatially periodic states are plotted. The light blue area corresponds to the region of bistability between $A_h$ and P$_1$, and spans the region between SN$_1^p$ and HH. The vertical dashed lines correspond to the diagrams shown in panel (a)-(c). Panels (1) and (2) show two examples of P$_1$ ($q=q_c$) and P$_2$ ($q=2q_c$) when $\Delta=1.5$. Adapted from \citet{parra-rivas_bifurcation_2018-1}.}
	\label{Fig_diapattern}
\end{figure} 

As mentioned previously, the formation of LPs and their bifurcation structure are directly related to the spatially periodic pattern arising from the HH bifurcation with wavenumber $q_c$. Therefore, it is essential to understand first the bifurcation features of such pattern states. For parameter values close to the HH point, periodic patterns are well described by the approximate asymptotic expression (\ref{Turing_pattern}). However, as the system parameters shift from HH, the accuracy of Eq.~(\ref{Turing_pattern}) diminishes. In this case, it is essential to use numerical path continuation algorithms to track the periodic solutions \citep{allgower_numerical_1990,doedel_numerical_1991,doedel_numerical_1991-1}. These methods, based on a predictor-corrector approach, permit the numerical tracking of a given state, here a spatially periodic state, as a function of a suitable control parameter. In the present case, the application of this technique leads to the bifurcation diagrams shown in Fig.~\ref{Fig_diapattern}(a)-(c), where the energy E$_1$ is defined as the $L^2$ norm of $A$,
\begin{equation}
{\rm E}_1\equiv||A||^2\equiv\frac{1}{L}\int_{-L/2}^{L/2}|A(x)|^2dx.
\end{equation}  
Figure~\ref{Fig_diapattern}(a) shows the bifurcation diagram for $\Delta=1.2$, corresponding to a cut of the $(\Delta,S)$-phase diagram shown in Fig.~\ref{Fig_diapattern}(d), where the main bifurcation curves of the system are plotted. For this value of $\Delta$, the primary periodic pattern P$_1$ arises from HH supercritically, and is therefore temporally stable [see the blue brach in Fig.~\ref{Fig_diapattern}(a)]. Increasing $S$, this state connects with a subsidiary primary pattern P$_2$ of wavenumber $2q_c$ originating at pb$_2$ and does so at a $2:1$ spatial resonance SR$_{2:1}$ (see close-up view). With increasing $S$, the P$_2$ pattern connects to a subsidiary primary pattern P$_4$ with $q=4q_c$ at a second SR$_{2:1}$ (not shown here). This process repeats, leading to a sequence of primary bifurcations pb$_{2^i}$ to patterns with wavenumber $q=2^iq_c$ ($i=0,1,2,\dots$) and associated SR$_{2:1}$ points, as described in \citet{parra-rivas_bifurcation_2018-1}. Unlike HH the subsidiary primary bifurcations cannot be characterized in terms of spatial dynamics. However, as recently shown by G\"artner {\it et al.}, they can be determined analytically in terms of transversality conditions \citep{gartner_bandwidth_2019}. 

Increasing $\Delta$ further, P$_1$ becomes subcritical at the degenerate HH point D$_1$ occurring at $\Delta=41/30$. This situation is shown in Fig.~\ref{Fig_diapattern}(b) for $\Delta=1.5$. In this case, P$_1$ is initially unstable but acquires stability in a saddle-node bifurcation SN$^p_1$. A representative example of this periodic state is shown in Fig.~\ref{Fig_diapattern}~(1). Thereafter P$_1$ remains stable all the way until it reaches SN$^p_2$, where it again loses stability, prior to connecting to P$_2$ [see profile shown in Fig.~\ref{Fig_diapattern}~(2)] at SR$_{2:1}$. Thus, in this regime there is a parameter interval where stable P$_1$ and $A_h$ coexist, an interval we refer to as the pattern-$A_h$ bistability region [see shaded box in Fig.~\ref{Fig_diapattern}(b)]. In the phase diagram shown in Fig.~\ref{Fig_diapattern}(d), the bistability region corresponds to the light blue area between SN$^p_1$ and HH.

\begin{figure}[!t]
	\centering\includegraphics[scale=0.9]{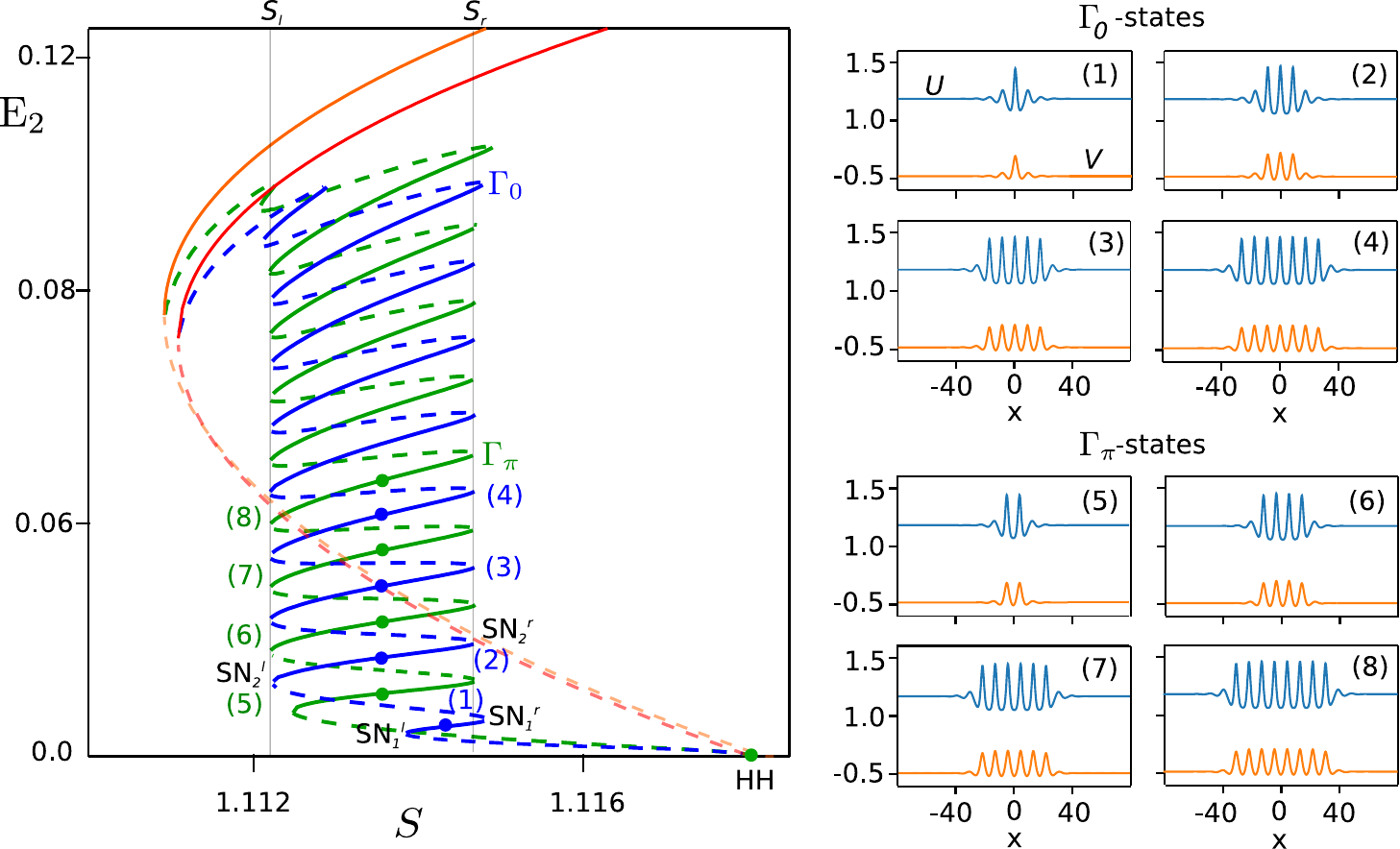}
	\caption{Localized patterns and the homoclinic snaking structure. We show the $L_2$ norm E$_2$ as a function of $S$ for $\Delta=1.5$. Solid (dashed) lines correspond to temporally stable (unstable) states. The blue snaking curve corresponds to the $\Gamma_0$ family of LPs. Panels (1)-(4) show some representative examples along this curve. The green snaking curve corresponds to $\Gamma_\pi$, and some representative LP examples along this curve are shown in panels (5)-(8). The snaking or pinning region is delimited by the $S_{l,r}$. The states $\Gamma_0$ and $\Gamma_\pi$ arise together in HH and connect with two periodic patterns of different wavelengths (see the red and orange curves). For both $\Gamma_{0,\pi}$ the saddle-node bifurcations are labeled SN$_i^{r,l}$ from the bottom to the top, with $i=1,2,3,\dots$ Adapted from \citet{parra-rivas_bifurcation_2018}.}
	\label{Hom_snaking}
\end{figure}

The bifurcation at pb$_2$ becomes degenerate at D$_2$ and beyond D$_2$ the pattern P$_2$ arises subcritically and stabilizes at SN$^p_3$. At this stage the bifurcation scenario is similar to that depicted in Fig.~\ref{Fig_diapattern}(c) for $\Delta=1.72$, where P$_1$ still connects with P$_2$ in SR$_{2:1}$, which now occurs very close to SN$^p_3$ [see the close-up view of Fig.~\ref{Fig_diapattern}(c)]. This bifurcation scenario persists for all $\Delta<2$. In the limit $\Delta\rightarrow2$ (i.e., when approaching the QZ point), the phase diagram of Fig.~\ref{Fig_diapattern}(d) shows how HH and the subsidiary bifurcation pb$_2$ tend asymptotically to SN$_b$, whereas SR$_{2:1}$ tends to SN$^p_3$. The diagrams shown in Fig.~\ref{Fig_diapattern}(a)-(c) reflect the bifurcations associated with P$_1$ and P$_2$ only, although similar transitions occur between P$_2$ and P$_4$, P$_4$ and P$_8$, and so on \citep{parra-rivas_bifurcation_2018-1}.
 
Furthermore, these patterns undergo a variety of other instabilities, such as Eckhaus and Hopf bifurcations, which have been analyzed in detail by different authors \citep{perinet_eckhaus_2017,delcey_periodic_2018,parra-rivas_bifurcation_2018-1,kholmyansky_optimal_2019-1}. For example, \citet{perinet_eckhaus_2017} and \citet{delcey_periodic_2018} perform an analytical study of the Eckhaus instability of supercritical patterns very close to HH. In highly nonlinear regimes, however, this approach is no longer valid and stability must be computed numerically as done in \citet{perinet_eckhaus_2017,parra-rivas_bifurcation_2018-1,kholmyansky_optimal_2019-1,gomila_observation_2020}.

We have focused here on the bifurcation structure of Turing patterns arising from HH, i.e., patterns with wavenumber $q_c$. However, the subsidiary patterns with wavenumber $q=2^iq_c$, $i=1,2,\dots$ that emerge from $A_h$ whenever $I_h>I_c$ undergo similar behavior, and other bifurcation structures organized through $3:1$ spatial resonances (SR$_{3:1}$) etc. are also present, as discussed further by \citet{perinet_eckhaus_2017}.

\subsection{Localized patterns and the snakes-and-ladders structure}
\label{sec:3.2}
The weakly nonlinear analysis carried out in Sec.~\ref{sec:2} revealed that whenever $41/30<\Delta<2$, weakly nonlinear LPs of the form (\ref{WNL_Wild_LP}) bifurcate subcritically from HH together with spatially periodic states of wavenumber $q_c$ [see Eq.~(\ref{Turing_pattern})]. Moreover, these LPs emerge in two families $\Gamma_0$ and $\Gamma_\pi$, corresponding to $\varphi=0$ and $\varphi=\pi$, respectively. Like the weakly nonlinear patterns, these asymptotic LP solutions are only valid very close to HH. However, the numerical path continuation methods applied in Sec.~\ref{sec:3.1} allow one to characterize such states in highly nonlinear regimes for parameters far from HH, and to compute their bifurcation diagrams. 

Figure~\ref{Hom_snaking} shows the resulting diagram computed for $\Delta=1.5$, where instead of E$_1$, we use the bifurcation measure E$_2\equiv||A-A_h||^2$ to better visualize the solution branches. The two families of solutions, $\Gamma_0$ and $\Gamma_\pi$, are plotted in blue and green, respectively. Both curves of solutions persist to finite amplitude and undergo {\it homoclinic snaking}: a sequence of back-and-forth oscillations in $S$ reflecting the successive nucleation of a pair of pattern peaks, one of each side of the structure, as one follows the diagram (i.e., $\Gamma_0$ and $\Gamma_\pi$) upwards. These oscillations occur within an interval $S_l<S<S_r$, known as the {\it snaking or pinning region}. The solution curves $\Gamma_0$ and $\Gamma_\pi$ undergo a sequence of saddle-node bifurcations SN$_i^{l,r}$ at which the LPs repeatedly gain and lose temporal stability. Some representative examples of these LPs are shown in Fig.~\ref{Hom_snaking}(1)-(8). The profiles shown in panels (1)-(4) belong to $\Gamma_0$, and consist of an odd number of pattern peaks embedded in an $A_h^b$ background. The solution profiles shown in panels (5)-(8) belong to $\Gamma_\pi$, and consist of an even number of pattern peaks embedded in $A_h^b$. The saddle-node bifurcations on either side of these curves converge exponentially and monotonically to the limits of the pinning region $S_l$ and $S_r$. 

In an infinite domain the peak nucleation process continues indefinitely. In a finite domain, however, this process must terminate, as the number of peaks allowed is constrained by the size of the domain. In periodic domains, like ours, $\Gamma_0$ and $\Gamma_\pi$ terminate near the saddle-node of one of the many subcritical periodic patterns emerging from $A_h$ for $I_h>I_c$ as shown in Fig.~\ref{Hom_snaking}.
  \begin{figure}[!t]
  	\centering\includegraphics[scale=0.9]{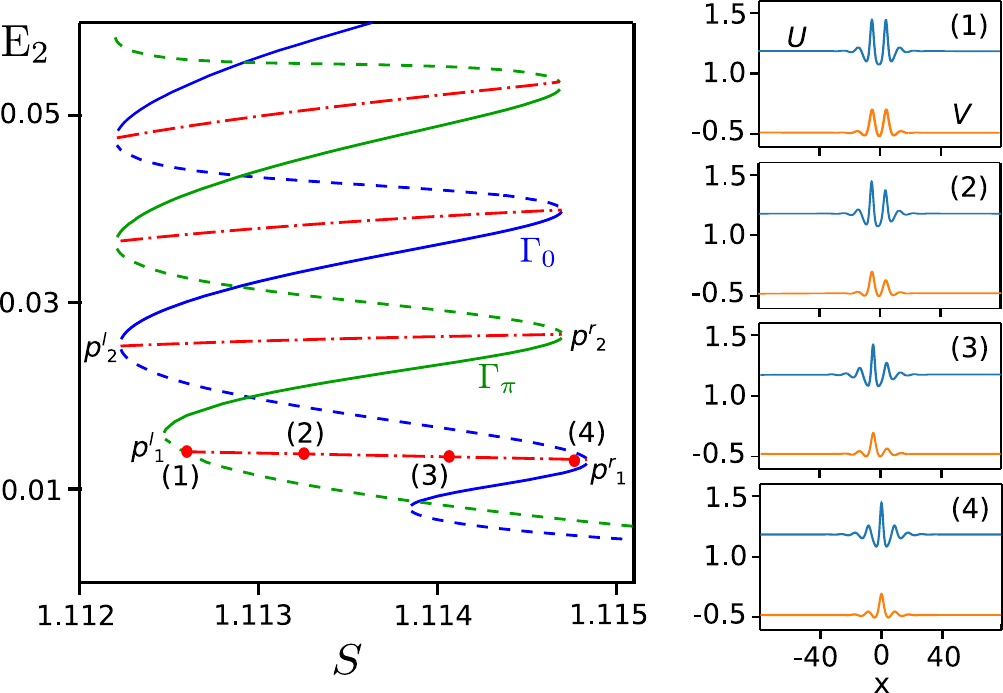}
  	\caption{The snakes-and-ladders bifurcation structure. The homoclinic snaking curves $\Gamma_0$ and $\Gamma_\pi$ are interconnected through a series of rung branches corresponding to asymmetric states. The states arise from symmetry-breaking pitchfork bifurcations labeled $p_i^{l,r}$ ($i=1,2,3,\dots$) occurring near the folds of the snaking curves. Labels (1)-(4) in the diagram correspond to the LP states shown on the right. Adapted from \citet{parra-rivas_bifurcation_2018}.}
  	\label{ladder}
  \end{figure}

Figure~\ref{ladder} shows a portion of the diagram shown in Fig.~\ref{Hom_snaking}, where we plot the rung states connecting $\Gamma_0$ and $\Gamma_\pi$. These branches correspond to traveling asymmetric states. These states move at a constant speed determined by the parameters and are temporally unstable. The rung states arise from secondary symmetry-breaking bifurcations ($p_i^{l,r}$) occurring near SN$_{i}^{l,r}$ on both $\Gamma_0$ and $\Gamma_\pi$. Some of these states are shown in Fig.~\ref{ladder}~(1)-(4). In Fig.~\ref{ladder}~(1), the two-peak profile bifurcating from $p_1^l$ is weakly asymmetric and therefore very similar to the completely symmetric state on the unstable $\Gamma_\pi$ branch. As $S$ increases, the peak on the right decreases in amplitude [see profiles (2) and (3)] until it rejoins $\Gamma_0$ in $p_1^r$, becoming the completely symmetric single peak LP shown in Fig.~\ref{ladder}~(4). The secondary bifurcations are pitchforks so each rung actually includes a pair of branches of asymmetric states with identical E$_2$, and related by the reversibility symmetry $A(x)\rightarrow A(-x)$. The rung states form, together with the homoclinic snaking, what is now known as a {\it snakes-and-ladders} structure, first identified by Burke and Knobloch in the context of the Swift-Hohenberg equation \citep{burkeknobloch2006,burke_snakes_2007}. 

As originally proposed by Woods and Champneys, the emergence of these LPs, and the homoclinic snaking that they undergo, is a consequence of a {\it heteroclinic tangle} present within $S_l<S<S_r$ arising from the transversal intersection of the unstable manifold of $A_h^b$ [$W^u(A_h^b)$] and the stable manifold of a given spatially periodic pattern P [$W^s(P)$] as $S$ varies and vice versa \citep{woods_heteroclinic_1999,beck_snakes_2009}. The first tangency between $W^u(A_h^b)$ and $W^s(P)$ at $S_l$ corresponds to the birth of Shilnikov-type homoclinic orbits bi-asymptotic to the bi-focus equilibrium $A_h^b$, while the last tangency at $S_r$ corresponds to their destruction. In fact, as revealed by Gomila {\it et al.}, the actual scenario in the context of Eq.~(\ref{LLE}) is rather more complex, and additional LPs and complexes arise from the heteroclinic tangle between the stable and unstable manifolds of the high amplitude stable pattern, and the unstable low amplitude one \citep{gomila_bifurcation_2007}. We refer to these works for a more detailed description of the heteroclinic tangle process.

\begin{figure}[!t]
\centering\includegraphics[scale=1]{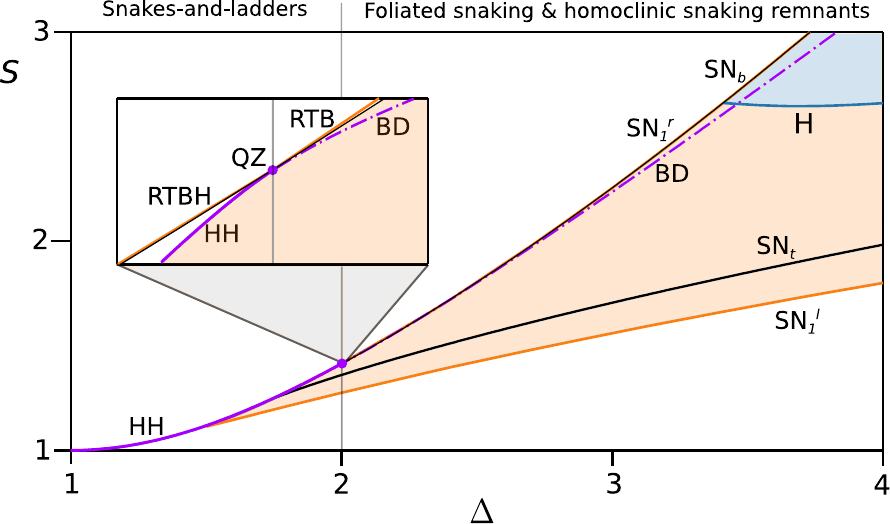}
\caption{Phase diagram in the $(\Delta,S)-$parameter space showing the main bifurcations of the system, and the region of existence and stability of the LSs. For $\Delta<2$ LPs exist between SN$_1^l$ and SN$_1^r$ and are organized within a snakes-and-ladders structure. For $\Delta>2$ the homoclinic snaking is destroyed but LPs persist for parameters $(\Delta,S)$ below BD. In contrast, spike LSs exist and undergo foliated snaking between SN$_1^l$ and SN$_1^r$. The region of existence of stable LSs is shown in light orange. For large values of $\Delta$ the system undergoes a Hopf (H) bifurcation and the spikes begin to oscillate, eventually leading to temporal and spatiotemporal chaotic dynamics (light blue area). The Hopf bifurcation arises from a Gavrilov-Guckenheimer (GG) codimension-two point occurring on SN$_1^r$. The inset shows a close-up view of the phase diagram around the QZ point from where all the spatial bifurcations arise. Adapted from \citet{parra-rivas_bifurcation_2018}. }
\label{Phase_anomalous}
\end{figure}

Thus far we have focused on a particular and representative detuning value: $\Delta=1.5$. However, the snakes-and-ladders structure persists within a larger parameter range extending to $\Delta=2$. The region of existence of this bifurcation structure can be computed by means of two-parameter continuation of the saddle-nodes SN$_i^{l,r}$ in the $(\Delta,S)$-parameter space. In doing so, we obtain the phase diagram shown in Fig.~\ref{Phase_anomalous}, where the main bifurcations of the system are plotted. The bifurcation curves include the HH (purple solid line), which becomes a BD transition for $\Delta>2$ (purple point-dashed line), the saddle-nodes of the homogeneous states SN$_{b,t}$, and the saddle-nodes of the single-peak LS SN$_1^{l,r}$.

When $\Delta>2$ the snakes-and-ladders structure is no longer present but spike LSs remain and are now organized in a new snaking structure called foliated snaking, as described next. On an infinite domain we expect that this structure extends only up to the BD line \citep{champneys_homoclinic_1998,parra-rivas_bifurcation_2018,verschueren_dissecting_2020} but this appears not to be the case on finite periodic domains where periodic arrays of spikes can be continued past the BD point into the region where the spike tails are all monotonic \citep{knobloch_stationary_2020}. This is a consequence of the fact that on such domains the global bifurcations (in space) that destroy these structures can no longer take place. This behavior is related to the scenarios shown in Figs.~\ref{unfoldingQZ}(b,c), and will be addressed in more detail in the next section.




\subsection{Foliated snaking and the remnants of the homoclinic snaking}\label{sec:3.3}
So far we have focused on the bifurcation structure of spatially periodic patterns and the localized patterns emerging from HH for $\Delta<2$. At $\Delta=2$ the system undergoes a QZ bifurcation resulting from the collision of the HH and the RTBH at SN$_b$. As a result, for $\Delta>2$, HH is replaced by a BD transition and $A_h^b$ is stable until SN$_b$, which now corresponds to a RTB spatial bifurcation. At this point, one may wonder what happens to the snakes-and-ladders bifurcation scenario, and whether LPs still exist or simply disappear. 


\begin{figure}[!t]
	\centering\includegraphics[scale=0.8]{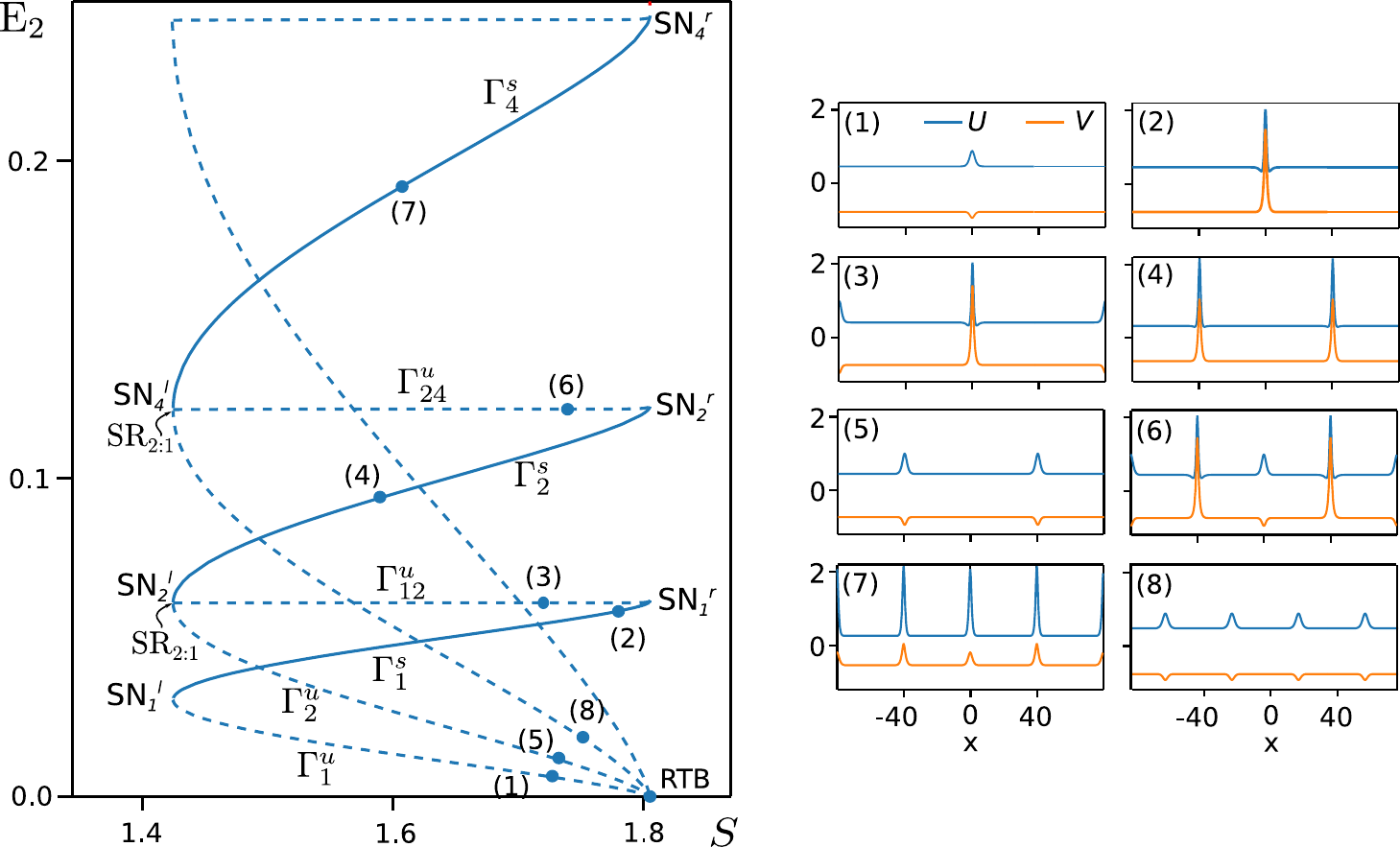}
	\caption{Bifurcation diagram showing the foliated snaking structure for $\Delta=2.5$. All the different unstable branches $\Gamma_{n}^u$, with $n=1,2,3,\dots$ emerge from SN$_b$ (RTB) and connect to one another through a sequence of $2:1$ spatial resonances SR$_{2:1}$ occurring nearby SN$_i^l$. Stable (unstable) branches are labeled with solid (dashed) lines. The blue dots on the foliated snaking branches correspond to the states shown in the panels on the right. Adapted from \citet{parra-rivas_bifurcation_2018}.}
	\label{foliated}
\end{figure}

Using the multiscale perturbation analysis of Sec.~\ref{sec:2}, one finds that whenever $\Delta>2$, weakly nonlinear tame LSs [see Eq.~(\ref{tame_LS})] bifurcate from SN$_b$ (i.e., a RTB). Although this solution is only valid near SN$_b$, numerical continuation of such states eventually leads to the bifurcation diagram shown in Fig.~\ref{foliated}. This bifurcation structure is known as {\it foliated snaking} and was first obtained by Ponedel and Knobloch in the context of periodically forced systems in space \citep{ponedel_forced_2016}. The small amplitude pulse emerging from SN$_b$ is like that shown in Fig.~\ref{foliated}~(1), and is associated with the solution branch $\Gamma_1^u$. Decreasing $S$ along $\Gamma_1^u$, this state grows in amplitude until it reaches SN$_1^l$, where it stabilizes and becomes the high amplitude {\it spike} shown in Fig.~\ref{foliated}~(2), corresponding to the solution branch $\Gamma_1^s$. Increasing $S$ further, this state eventually undergoes a saddle-node bifurcation (SN$_1^r$) where it loses stability. Soon after SN$_1^r$ is passed, a new small amplitude spike is nucleated at a separation $L/2$ from the high amplitude peak as shown in Fig.~\ref{foliated}~(3). We label the corresponding solution branch $\Gamma_{12}^u$. The newly nucleated spike grows in amplitude as $S$ decreases until it becomes identical to the original spike. This occurs at a $2:1$ spatial resonance SR$_{2:1}$, where $\Gamma_{12}^u$ connects with $\Gamma_2^u$ and $\Gamma_2^s$, very close to SN$_2^l$. Along $\Gamma_2^s$, the two spikes grow together as $S$ increases [see Fig.~\ref{foliated}~(4)], while the opposite occurs along $\Gamma_2^u$ [see profile Fig.~\ref{foliated}~(5)] as the amplitude of the two-peak state decreases to zero at SN$_b$ and the branch connects to $A_h^b$. Beyond SN$_2^r$ (see $\Gamma_{24}^u$), intermediate spikes nucleate midway between the large spikes already present [see Fig.~\ref{foliated}~(6)], and these grow to full amplitude by the time they reach the next SR$_{2:1}$ point near SN$_4^l$ where $\Gamma_{4}^u$ connects to $\Gamma_{4}^s$. Two characteristic states from these branches are shown in Fig.~\ref{foliated}~(7) and Fig.~\ref{foliated}~(8). The very same process repeats, resulting in a cascade of equally spaced states with $2^n$ spikes.

The foliated snaking scenario resembles the bifurcation structure associated with the periodic patterns discussed in Sec.~\ref{sec:3.1}. Indeed, the diagram shown in Fig.~\ref{foliated} is very similar to that plotted in Fig.~\ref{Fig_diapattern}(c) once the background field $A_h^b$ is removed from the first one. A first explanation of this similarity can be found in the spatial dynamics analysis carried out in Sec.~\ref{sec:2}. Let us imagine a periodic pattern bifurcating from HH with $q=q_c$ in the regime $\Delta<2$. When the system approaches QZ from below ($\Delta\rightarrow 2^{-}$), $q_c\rightarrow 0$, and in a finite system, a spatially periodic pattern with domain-size wavelength becomes indistinguishable from the spike shown in Fig.~\ref{foliated}~(2). This transition indicates that at the QZ point a global homoclinic bifurcation takes place where the pattern wavelength $2\pi/q_c$ diverges, and the spatially periodic state (a limit cycle) becomes a spike (a tame homoclinic orbit). This is the situation described schematically in Fig.~\ref{unfoldingQZ}(b). Thus, in this limit, P$_1$ becomes a single spike, P$_2$ two equidistant spikes, and so on. This new configuration persists for $\Delta>2$ in such a way that the foliated snaking preserves the bifurcation structure of the $q_c$-patterns. Furthermore, on top of this bifurcation structure, similar foliated snaking structures can be found for states with $n\in\mathbb{N}^+$ equidistant peaks, where the different branches also emerge from SN$_b$ but are now interconnected through $n:1$ spatial resonances SR$_{n:1}$ \citep{parra-rivas_bifurcation_2018}. 

\begin{figure}[!t]
	\centering\includegraphics[scale=0.9]{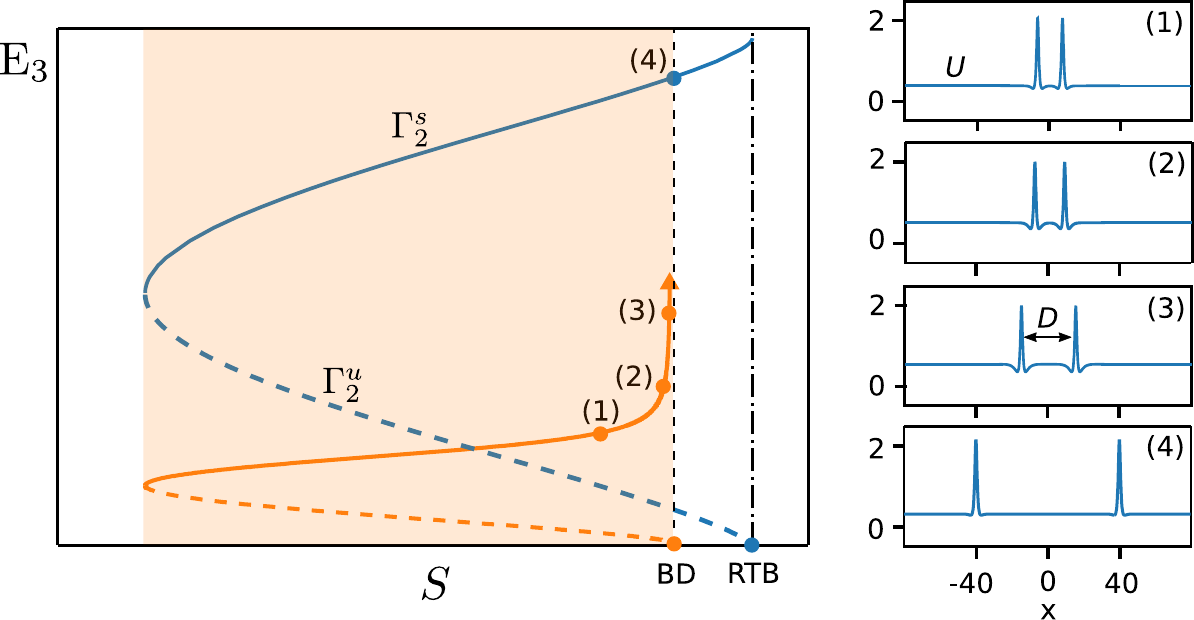}
	\caption{Reconnection of the remnants of the homoclinic snaking branches with the foliated snaking for $\Delta>2$. This diagram shows, through the bifurcation measure E$_3$, two branches of foliated snaking, $\Gamma_2^u$ and $\Gamma_2^s$ (blue lines), corresponding to the two equally-spaced spikes shown in panel (4). The orange curve shows the remnants of two homoclinic snaking branches corresponding to two-peak $\Gamma_\pi$ LPs. While foliated snaking arises from the RTB-SN$_b$ (see vertical point-dashed line), the LP branches emerge from a global homoclinic bifurcation occurring at the BD transition (see vertical dashed line). Approaching BD, the LP peak separation $D$ grows drastically until it reaches the maximum separation $L/2$ exactly at the BD point. Panels (1)-(3) show the change in the LP profiles along this curve. At the BD transition the LPs becomes the state shown in panel (4).  Adapted from \citet{parra-rivas_bifurcation_2018}. }
	\label{remanats}
\end{figure}

Foliated snaking organizes periodic patterns or patterns of equally spaced spikes, but it does not reveal the existence and potential organization of the LPs discussed in Sec.~\ref{sec:3.1}. At the QZ point, homoclinic snaking is destroyed, and for $\Delta>2$, LPs are organized differently \citep{parra-rivas_bifurcation_2018}. To reveal this bifurcation scenario, one can path-continue any LP in the two parameters $(\Delta,S)$ from $\Delta<2$ to $\Delta>2$, and after that compute the solution branches as a function of $S$ for a fixed value of $\Delta$. The result of this computational approach is shown schematically in Fig.~\ref{remanats}. Here, in order to better visualize the different solution branches, we defined a new bifurcation measure ${\rm E}_3\equiv{\rm E}_2\cdot D$, with $D$ the separation between peaks in the LSs. The blue lines correspond to the $\Gamma_2^{u,s}$ solution branches of the foliated snaking associated with two identical equally spaced spikes [see Fig.~\ref{remanats}~(4)]. The orange lines show two branches of the two-peak LP homotopically related with $\Gamma_\pi$ [see Fig.~\ref{Hom_snaking}(b)]. The point-dashed vertical line in Fig.~\ref{remanats} marks the location of SN$_b$ (i.e., the RTB bifurcation) where the foliated snaking emerges, while the vertical dashed line marks the position of the BD transition. LP branches do not bifurcate from the RTB, in contrast to the foliated snaking, but they finish very close to BD. Indeed, the use of E$_3$ reveals that $D$ diverges when approaching BD, as can also be seen in the profiles shown in panels Fig.~\ref{remanats}~(1)-(3). This scenario corresponds to that shown in Fig.~\ref{unfoldingQZ}(c). Below BD, LPs still form through a {\it heteroclinic tangle} as described in Sec.~\ref{sec:3.2}, and remnants of the homoclinic snaking branches can be found. The divergence in $D$ undergone by the LPs as $S\rightarrow S_{\rm BD}$ corresponds to the divergence of the wavelength of the periodic pattern involved in the tangle, and it reveals the occurrence of a global homoclinic bifurcation at $S_{\rm BD}$. Indeed, in this homoclinic bifurcation, periodic pattern states (limit cycles) turn into spikes (tame homoclinic orbits). This global phenomenon appears in different contexts, and is referred to as {\it a blue sky catastrophe} \citep{devaney_blue_1977}, {\it a wavelength blow-up} \citep{vanderbauwhede_homoclinic_1992}, or {\it a broom} global bifurcation \citep{verschueren_model_2017}. The mathematical theory describing the system dynamics close to this global bifurcation has been recently developed by \citet{verschueren_dissecting_2020}, and the same transition has been identified in other systems \citep{verschueren_model_2017,verschueren_dissecting_2020,knobloch_stationary_2020}.

\section{Localized structures in the normal regime: Collapsed snaking}\label{sec:4}

In the normal dispersion regime ($\nu=-1$), the emergence of LSs is related to the coexistence of two different HSSs. In this regime, the stability and configuration of the different spatial bifurcations undergone by $A_h$ is very different from the anomalous regime, as shown in Fig.~\ref{HSS_unfolding}(e)-(h). In the monostable regime ($\Delta<\sqrt{3}$), $A_h$ is always stable, and no LSs exist. For $\Delta>\sqrt{3}$, however, the coexistence between $A_h^b$ and $A_h^t$ allows for the formation of DWs (i.e., heteroclinic orbits) connecting them. Single DWs drift with a constant speed, which depends on the control parameters of the system. However, at the Maxwell point of the system, this speed vanishes. Close to this Maxwell point, DWs can interact and lock to one another, thus also leading to zero speed. In this way, dark LSs of different widths can be formed. The resulting LSs are organized in a particular bifurcation structure known as {\it collapsed snaking}, whose morphology is a direct consequence of the DW interaction and locking \citep{knobloch_homoclinic_2005,yochelis_reciprocal_2006}. 

The formation of LSs in the normal regime has been addressed in a number of theoretical works \citep{godey_stability_2014-1,lobanov_frequency_2015,parra-rivas_origin_2016,parra-rivas_dark_2016,gartner_bandwidth_2019}, and their existence has been confirmed experimentally in microresonators \citep{xue_mode-locked_2015,nazemosadat_switching_2019} and pulse-pumped fiber cavities \citep{garbin_experimental_2017}. In this section we present the main results regarding the origin and bifurcation structure of dark LSs in this regime of operation.  

\subsection{Dark localized states and the collapsed snaking diagram}\label{sec:4.1}
\begin{figure}[!t]
	\centering\includegraphics[scale=0.85]{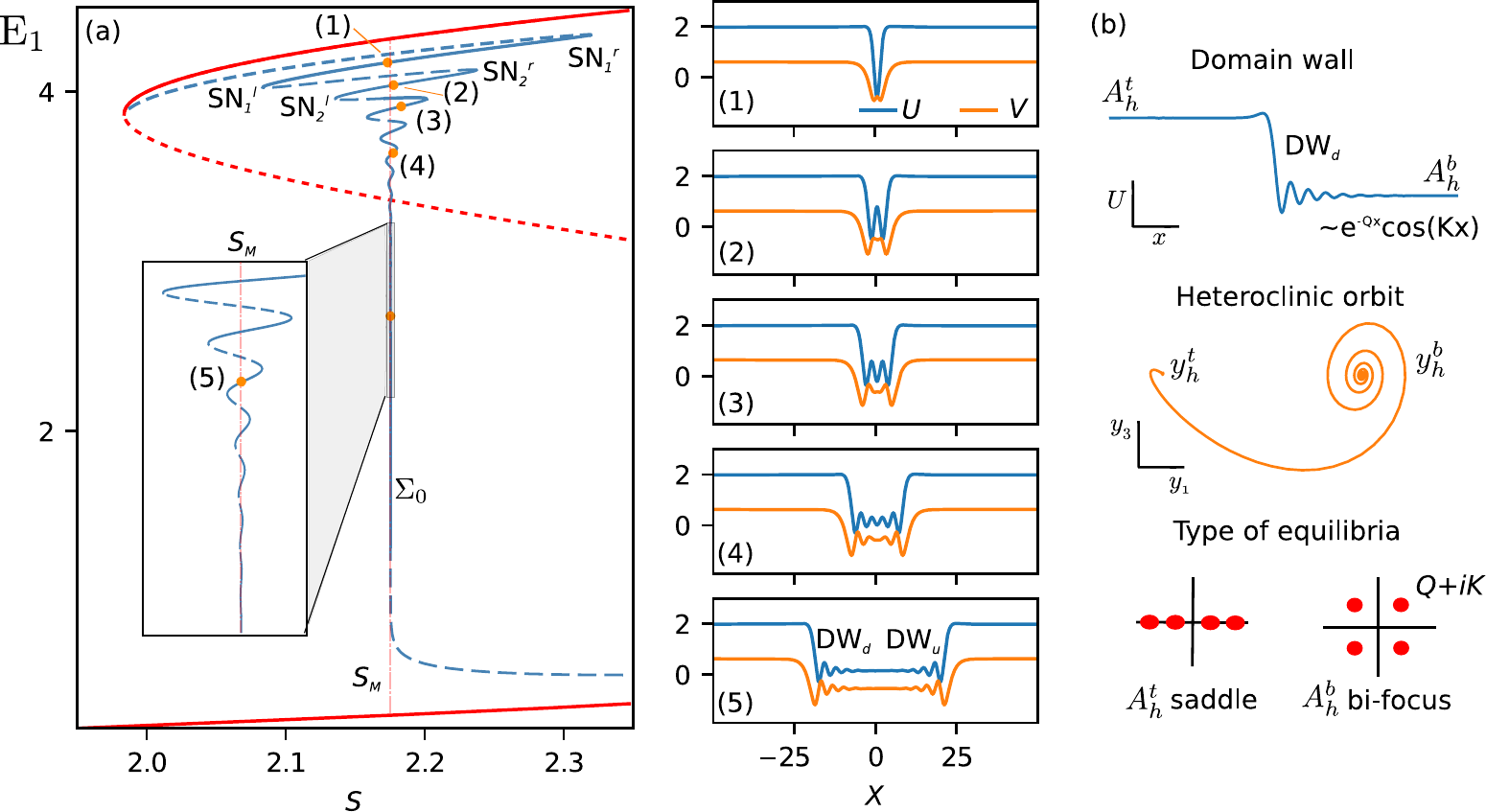}
	\caption{Collapsed snaking bifurcation structure. Panel (a) shows the bifurcation structure for $\Delta=4$, where E$_1$ is plotted as a function of $S$. The HSS $A_h$ is shown in red, while the blue curve is the collapsed snaking branch $\Sigma_0$. Solid (dashed) lines represent stable (unstable) LS branches. The labels (1)-(5) on the stable branches correspond to the LSs shown to the right. The close-up view in the main panel shows that the collapsed snaking behavior persists asymptotically close to the Maxwell point $S_M$. Panel (b) shows the morphology of DW$_d$ corresponding to panel (5). The tails of the DW are defined by the spatial eigenvalues associated with the equilibria $A_h^{b,t}$. Adapted from \citet{parra-rivas_dark_2016}.}
	\label{collapsed1}
\end{figure}

Figure~\ref{collapsed1} shows an example of collapsed snaking for $\Delta=4$, where E$_1$ is plotted as a function of $S$. The solution branches in red are those corresponding to $A_h$, whereas those associated with the LSs are shown in blue.
In the range of parameters shown in the diagram, $A_h^t$  ($A_h^b$) remains stable all the way until SN$_t$ (SN$_b$). In spatial dynamics terms, SN$_t$ corresponds to a RTB bifurcation, and weakly nonlinear states emerge from it in the form of tame homoclinic orbits (see Eq.~(\ref{tame_LS}) in Sec.~\ref{sec:2.2}). Numerical continuation of these solutions to parameter values far from SN$_t$ yields the blue solution curve $\Sigma$ shown in Fig.~\ref{collapsed1}. $\Sigma$ experiences a sequence of damped back-and-forth oscillations in $S$ around the Maxwell point of the system, $S=S_M$, and eventually collapses onto it. The morphology of this snaking curve is very different from the standard homoclinic snaking depicted in Sec.~\ref{sec:3}, which is why this diagram is known as {\it collapsed snaking} \citep{knobloch_homoclinic_2005,yochelis_reciprocal_2006}. Some representative examples of dark LSs along $\Sigma$ are shown in Fig.~\ref{collapsed1}~(1)-(5).

Let us briefly discuss how these states arise and change all along the diagram. The weakly nonlinear LSs first arise as unstable small amplitude holes in $A_h^t$ and bifurcate from SN$_t$. Following this unstable branch towards higher values of $S$, the amplitude of the LSs increases, and eventually the branch $\Sigma$ undergoes a first saddle-node bifurcation SN$_1^r$, where the LSs stabilize, and they remain stable until SN$_1^l$. At this stage, the LS resembles that depicted in panel (1) of Fig.~\ref{collapsed1}.  Soon after passing SN$_1^l$, the nucleation of a spatial oscillation (SO) around $x=0$ takes place, such that the inner part of the LS is filled with a portion of $A_h^b$. An example of this new state, once SN$_2^r$ is passed, is shown in Fig.~\ref{collapsed1}~(2). The SO nucleation process continues with decreasing E$_1$, leading to the sequence of LSs shown in panels (3)-(5). As a consequence, the LSs widen as $\Sigma$ asymptotically approaches the Maxwell point $S_M$. Figure~\ref{collapsed1}~(5) shows a LS close to $S_M$. At this stage, one can easily identify two well-formed DWs, namely DW$_u$ and DW$_d$, connecting $A_h^b$ and $A_h^t$. A close-up view of DW$_d$ is shown in Fig.~\ref{collapsed1}(b) together with the corresponding heteroclinic orbit. In terms of spatial dynamics, $A_h^t$ is a saddle equilibrium, whereas $A_h^b$ is a bi-focus, as shown by the spatial eigenvalues in Fig.~\ref{collapsed1}(b). Thus, the heteroclinic orbit leaves  $A_h^t$ monotonically, but it approaches $A_h^b$ in a damped oscillatory fashion. The part of the DW that approaches $A_h^b$ in this manner is typically called an {\it oscillatory tail}.

Decreasing E$_1$ further [see the bottom part of Fig.~\ref{collapsed1}], the branch $\Sigma$ eventually separates from $S_M$, and continues to $A_h^b$, where it disappears close to HH as described below.
\begin{figure}[!t]
	\centering\includegraphics[scale=1]{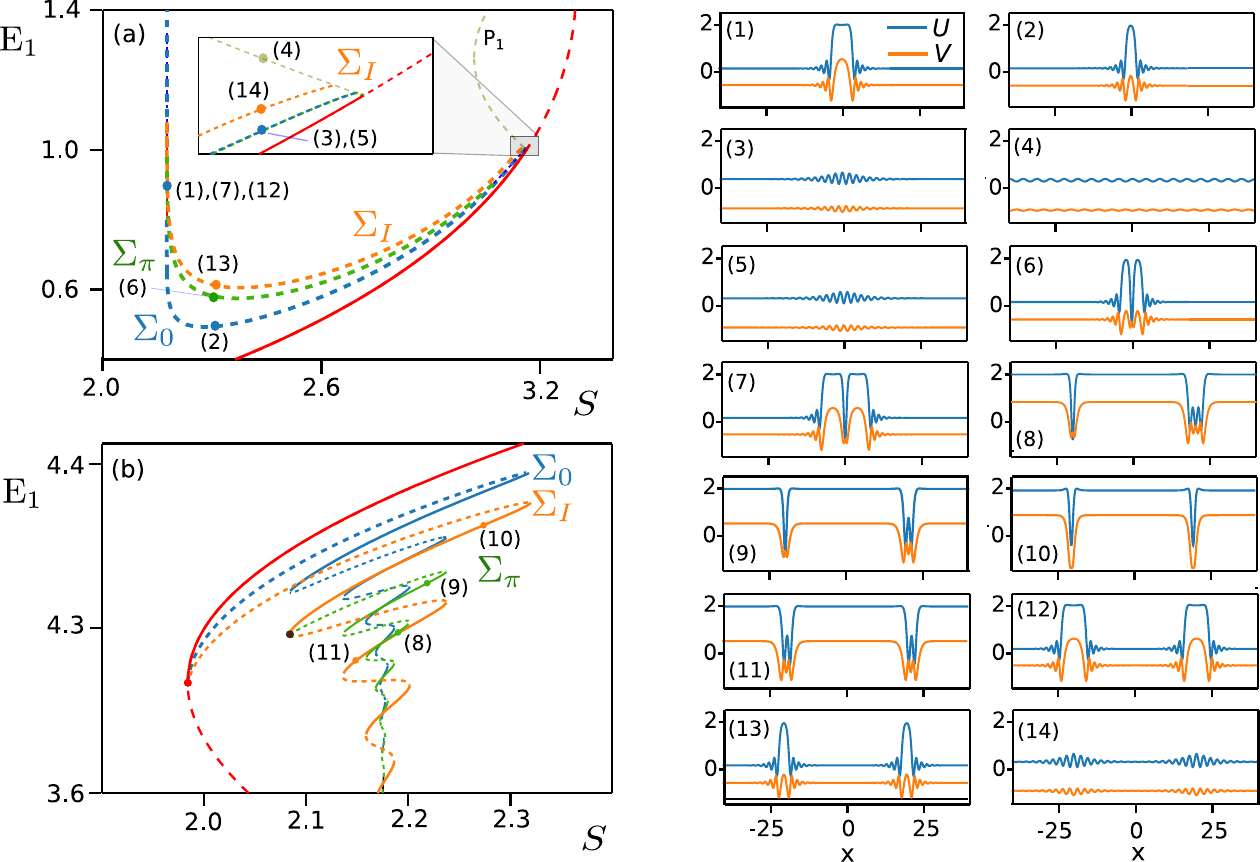}
	\caption{Panel (a) shows the reconnection of the collapsed snaking branch $\Sigma_0$, plotted in Fig.~\ref{collapsed1}, with $A_h$ near HH. The solution curves $\Sigma_\pi$ and $\Sigma_I$ also arise near HH and also undergo collapsed snaking. Panel (b) shows the top part of the collapsed snaking branch $\Sigma_0$ shown in Fig.~\ref{collapsed1}, together with $\Sigma_\pi$ and $\Sigma_I$. The labels (1)-(14) correspond to the states shown on  the right.  Adapted from \citet{parra-rivas_dark_2016}.}
	\label{collapsed2}
\end{figure}
This situation is shown in detail in Fig.~\ref{collapsed2}(a). In view of the periodic boundary conditions, when DW$_u$ and DW$_d$ move apart from $x=0$ they also approach one another at $x=L/2$ albeit back-to-back. In this context, the resulting state is a bright LS with a SO like that shown in Fig.~\ref{collapsed2}~(1), once a translation by $L/2$ has been taken into account. Thus as $E_1$ decreases the dark LS turn into bright LS. Increasing $S$ further, the bright state becomes that shown in Fig.~\ref{collapsed2}~(2), and very close to HH it reduces to the asymptotic LS calculated in Sec.~\ref{sec:2.1} for $\varphi=0$. In the following, we rename $\Sigma$ as $\Sigma_0$. Due to finite domain size effects, $\Sigma_0$ does not terminate exactly at HH, but at a subcritical pattern [see Fig.~\ref{collapsed2}~(4)] emerging from it.

\subsection{Secondary solution branches}
Apart from the dark/bright LSs belonging to $\Sigma_0$, there are other families of solutions which are interconnected. Close to HH, the asymptotic analysis carried out in Sec.~\ref{sec:2} shows that there must be another family of states characterized by $\varphi=\pi$ ($\Sigma_\pi$). Like $\Sigma_0$ this curve arises subcritically from the periodic pattern state, as shown in the close-up view of Fig.~\ref{collapsed2}(a). At this stage, a $\Sigma_\pi$ state resembles that depicted in panel Fig.~\ref{collapsed2}~(5), and possesses a minimum at $x=0$. Moving away from HH, the two central peaks grow [see profile in Fig.~\ref{collapsed2}~(6)] until their amplitude reaches $A_h^t$ [Fig.~\ref{collapsed2}~(7)]. The top of the two peaks then flattens forming two plateaus around  $A_h^t$, separated by a dark spike (hole) at $x=0$. During the flattening process, two DWs form that move farther and farther from $x=0$ as one proceeds up  $\Sigma_\pi$. Eventually, this state resembles a composition of two separate states: a dark spike and a dark LS with several SOs [Fig.~\ref{collapsed2}~(8)]. During this process, $\Sigma_\pi$ snakes around $S_M$ as depicted in Fig.~\ref{collapsed2}(b). While following the $\Sigma_\pi$ branch upwards, the SOs progressively disappear eventually resulting in a state formed of two identical dark spikes. When this occurs, $\Sigma_\pi$ collides with another collapsed snaking curve $\Sigma_I$ [see the orange curve in Fig.~\ref{collapsed2}(b)], corresponding to a pair of identical LS like that shown in Fig.~\ref{collapsed2}~(10)-(12). Increasing E$_1$, $\Sigma_I$ connects back to $A_h$ at SN$_t$ but proceeding down in $\Sigma_I$, each of the constituent LS behaves the same way, and the branch eventually connects back to the periodic pattern, but at a larger amplitude since $\Sigma_I$ in effect represents a single pulse state on the half-domain [see Fig.~\ref{collapsed2}(a)].
\begin{figure}[!t]
	\centering\includegraphics[scale=0.8]{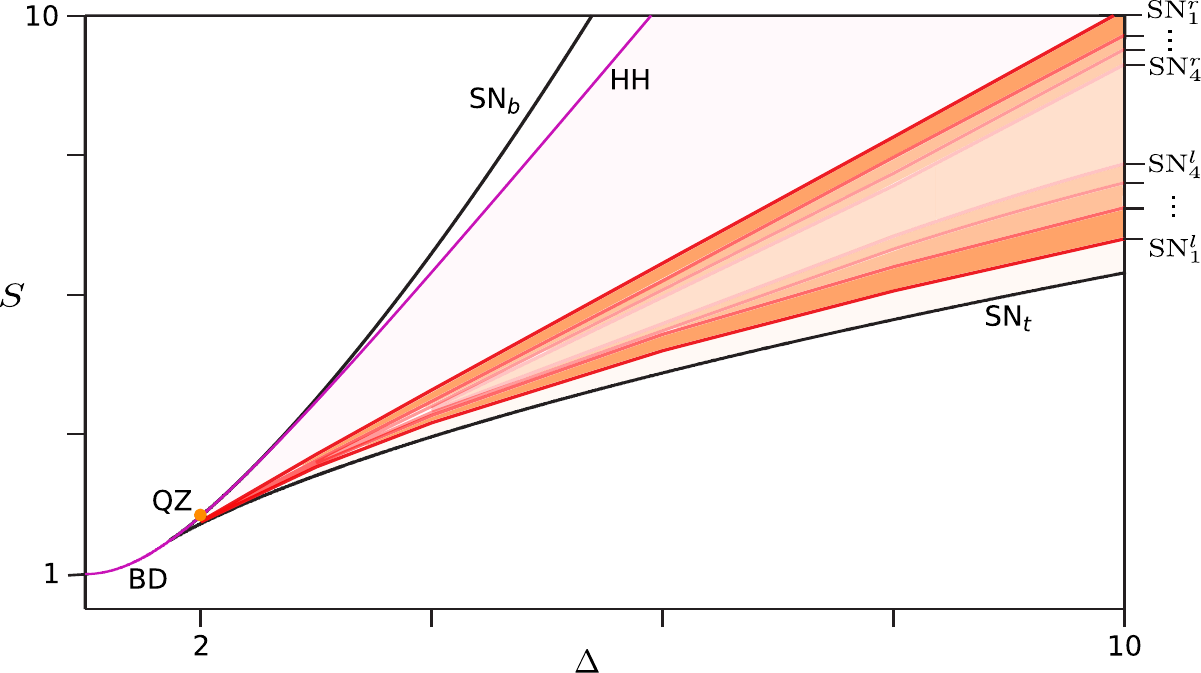}
	\caption{Phase diagram in the $(\Delta,S)-$parameter space for the normal regime. The bifurcation lines plotted here are: the HH-BD line, the saddle-nodes of the $A_h$, SN$_{b,t}$, and a number of saddle-node bifurcations SN$_i^{l,r}$ ($i=1,2,3,\dots$) associated with the collapsed snaking branch $\Sigma_0$ shown in Fig.~\ref{collapsed2}. The graded orange areas show the regions of existence of the different types of dark LSs. Adapted from \citet{parra-rivas_dark_2016}.}
	\label{collapsed3}
\end{figure}

Thus far we have focused on a particular value of the detuning, namely $\Delta=4$. One may wonder if this type of LS and the collapsed snaking associated with them, persists to other parameter regimes. This question is answered through a two-parameter continuation (in $\Delta$ and $S$) of the different saddle-node bifurcations SN$_i^{l,r}$ of $\Sigma_0$ (Fig.~\ref{collapsed3}). For increasing values of $\Delta$, the region of bistability between $A_h^t$ and $A_h^b$ expands, and so does the region of existence of dark LSs. When $\Delta$ decreases, however, both regions shrink and the different saddle-node bifurcations (SN$_i^{l,r}$, for $i=1,2,3,\dots$) collide in a cascade of cusp bifurcations, such that the widest LSs disappear first, while the single dark spike is the last one to disappear.

In the range of parameters examined here, the periodic pattern that arises from HH is unstable. Therefore it does not play any role in the formation of LSs. Nevertheless, the periodic pattern might be stabilized at much larger values of $\Delta$, which could lead to a very rich scenario involving tristability between the pattern, $A_h^t$ and $A_h^b$. In such a situation, a new type of hybrid LSs may arise as has been shown in other contexts \citep{zelnik_implications_2018}. The exploration of such a potential scenario requires further investigation.


\subsection{Domain wall locking as a mechanism to form localized structures}
The emergence of dark LSs, and their organization in a collapsed snaking structure, can be understood from a physical perspective based on the interaction and locking of DWs \citep{coullet_nature_1987}. In our system, DWs form between the two non-equivalent HSSs $A_h^t$ and $A_h^b$, and they drift at constant speed depending on the control parameters \citep{chomaz_absolute_1992}. At the Maxwell point $S_M$, the speed is zero and DWs are stationary. Close to that point, the interaction of two DWs with different polarity, DW$_u$ and DW$_d$, and separated by a distance $D$, can be phenomenologically described by the equation
\begin{equation}\label{locking}
\partial_t D=\rho e^{-|Q|D}{\rm cos}(KD)+\eta\equiv f(D),
\end{equation}  
where $Q$ and $K$ correspond to the real and imaginary parts of the spatial eigenvalue associated with $A_h^b$ (and so is responsible for the oscillatory tail of the DW), $\varrho$ depends on the parameters of the system, and $\eta$ is proportional to the distance from the Maxwell point (i.e., $\eta\sim S-S_M$). Equation~(\ref{locking}) cannot be explicitly derived from the LL equation (\ref{LLE}) for a number of reasons (e.g., the absence of an analytical DW solution), and we present it here simply to illustrate the locking mechanism of DWs. However, reduced equations such as Eq.~(\ref{locking}) can be explicitly derived in other contexts \citep{coullet_nature_1987,coullet_localized_2002,clerc_analytical_2010,escaff_localized_2015}, and a perturbed version of Eq.~(\ref{locking}) can be found close to the onset of nascent bistability (i.e., close to $\Delta=\sqrt{3}$) \citep{tlidi_localized_2015,clerc_nonlocal_2020-1}. 
 \begin{figure}[!t]
 	\centering\includegraphics[scale=1]{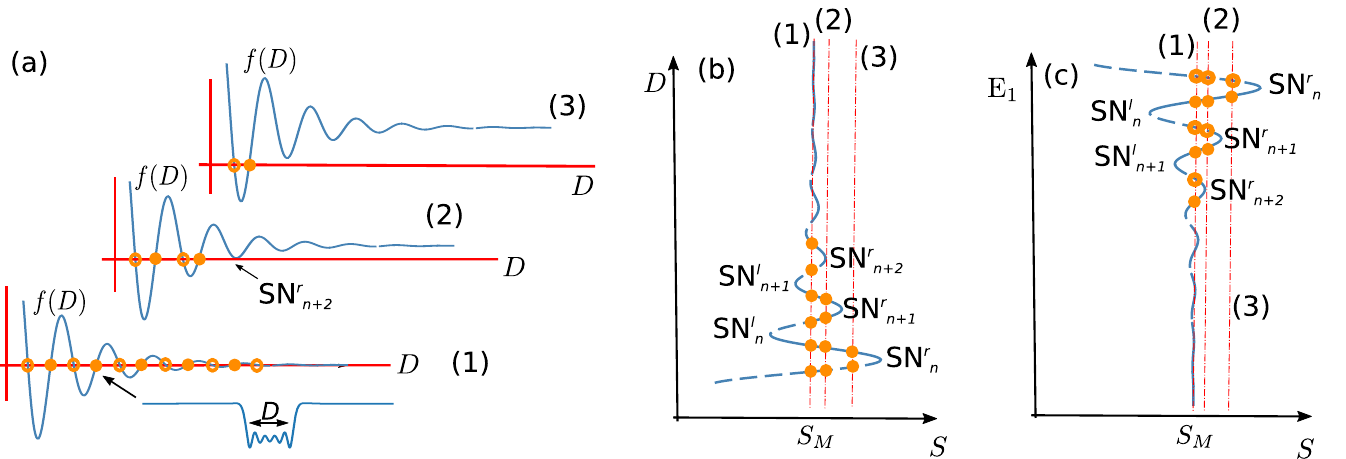}
 	\caption{Schematic representation of the interaction and locking of domain walls. Panel (a) shows the equilibria of Eq.~(\ref{locking}) for different values of $\eta$: $\eta=0$ in (1), and $\eta_{1,2}$, with $\eta_2>\eta_1$ in (2) and (3). Stable (unstable) states are labeled with full $\bullet$ (empty $\circ$) circles. Panel (b) shows the bifurcation diagram arising from the locking mechanism shown in (a), where points on the stable (unstable) parts of the diagram correspond to LSs with $D$ equal to the stable (unstable) zeros of (a) for each value of $S$. Panel (c) represents the information shown in panel (b) but using E$_1$ instead of $D$. This last diagram is a schematic picture of the collapsed snaking shown in Fig.~\ref{collapsed1}.}
 	\label{esquema_collapsed}
 \end{figure}
 Figure~\ref{esquema_collapsed}(a) shows $f(D)$ for three different values of $\eta$ (see the blue curves). The intersections of the curves with the horizontal axis correspond to the equilibria of Eq.~(\ref{locking}), and therefore to the locking of DWs and subsequent formation of LSs. In all cases stable and unstable separations (marked using $\bullet$ or $\circ$ respectively) alternate in $D$.

At $S=S_M$ [see Fig.~\ref{esquema_collapsed}(a)~(1)], the locking separations can be derived analytically and read $D_n^0=(2n+1)K/2\pi$. Each of these separations can be mapped to the bifurcation diagram shown in Fig.~\ref{esquema_collapsed}(b), where $D$ is plotted as a function of $S$, and that sketched in Fig.~\ref{esquema_collapsed}(c), which shows E$_1$ as a function of $S$. In both diagrams, stable (unstable) solution branches are plotted using solid (dashed) lines. Panels (2) and (3) in Fig.~\ref{esquema_collapsed}(a) show the situation for two different values of $S$ slightly above $S_M$. In panel (2) four intersections, two stable ones and two unstable ones, are shown, which map to the vertical dashed line (2) plotted in Fig.~\ref{esquema_collapsed}(b) and (c). With this mapping, the tangency shown in Fig.~\ref{esquema_collapsed}(a) corresponds to the saddle-node bifurcations shown in panels (b) and (c). Increasing $\eta$ a bit more [see panel (3) in Fig.~\ref{esquema_collapsed}(a)] leaves only one pair of stable/unstable equilibria, which defines the two points on the vertical dashed line (3) in Figs.~\ref{esquema_collapsed}(b) and (c). Hence, the change in the locking separations in Fig.~\ref{esquema_collapsed}(a) arising from shifting the blue curves upwards or downward (i.e., by changing $S$) determines the collapsed snaking curves shown in Figs.~\ref{esquema_collapsed}(b) and (c).

\section{Oscillatory and chaotic dynamics}\label{sec:5}
The static LSs described in the previous sections can also exhibit very rich dynamical behavior such as temporal oscillations, also known as {\it breathing behavior}, temporal chaos, and spatiotemporal chaos, which has been studied by many authors both experimentally and theoretically. In this section we briefly discuss, from a bifurcation perspective, some of the main features of the resulting dynamics in the anomalous and normal regimes, and refer to the original work for more details.

\begin{figure}[!t]
	\centering\includegraphics[scale=0.8]{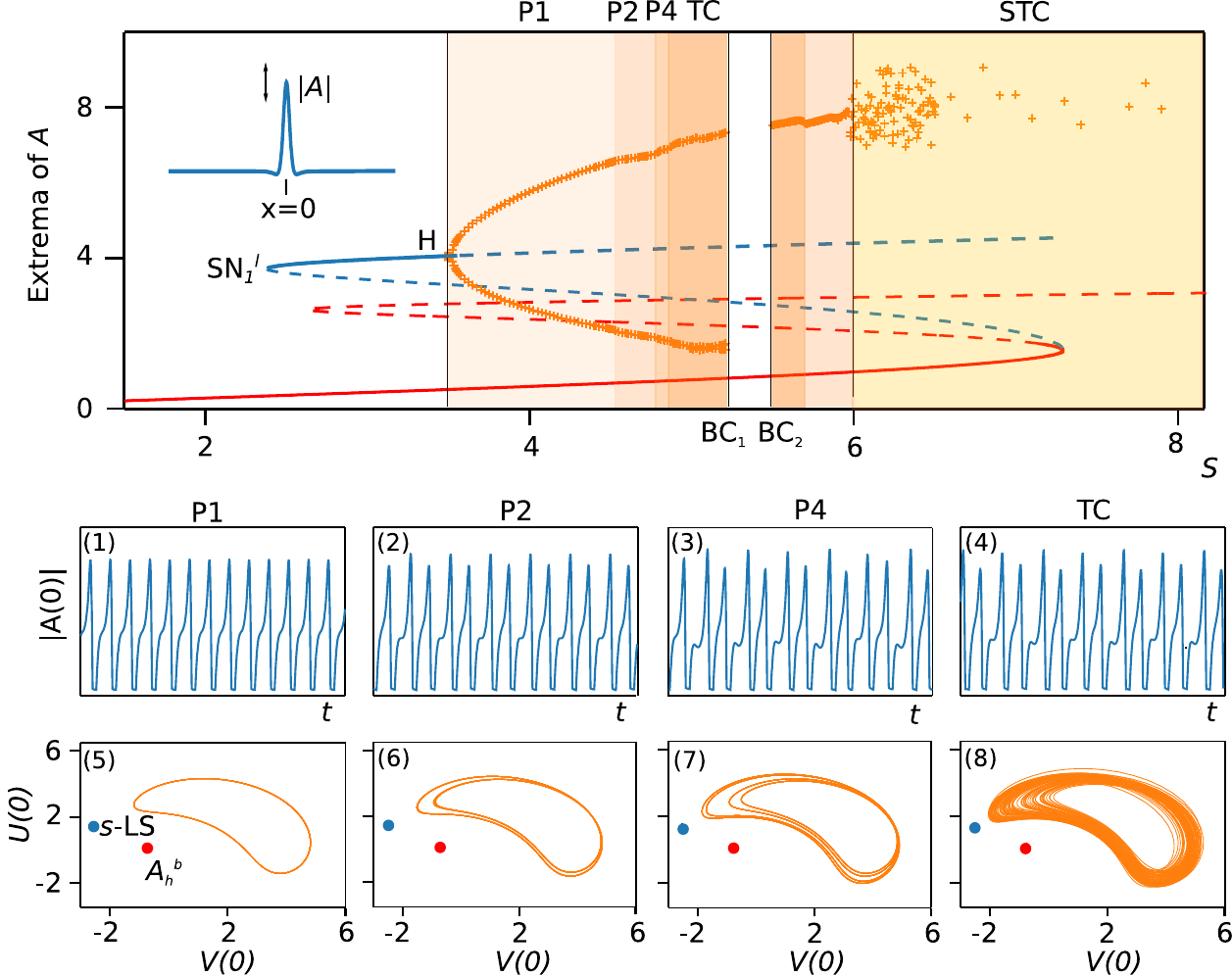}
	\caption{Bifurcation diagram and oscillatory behavior in the anomalous regime for $\Delta=7$. This diagram shows the HSS $\sqrt{I_h}$ in red, the peak value of the spike intensity $|A(0)|$ (blue curve), and the maxima and minima of the oscillation amplitude intensity at $x=0$. The different colored areas correspond to different types of oscillatory dynamics: period-1 oscillations (P1), period-2 (P2), period-4 (P4), temporal chaos (TC) and spatiotemporal chaos (STC). The letter H denotes a supercritical Hopf bifurcation while BC$_i$ correspond to boundary crises of the chaotic attractor. Panels (1)-(4) show the time series of the spike at its center $x=0$, and panels (5)-(8) show the two-dimensional phase space obtained from the projection of the oscillatory dynamics on the variables $U(0)$ and $V(0)$.}
	\label{breathers1}
\end{figure}

\subsection{Breathers in the anomalous regime}
In the anomalous regime, oscillatory and chaotic behavior was first identified experimentally in a series of seminal works in the context of fiber cavities \citep{leo_dynamics_2013}, and later in microresonators \citep{lucas_breathing_2017,yu_breather_2017}. In these papers, a breather consists in a bright spike whose amplitude oscillates in time with a fixed period, while preserving its position. The dynamics of dispersive Kerr breathers have been analyzed theoretically within the framework of Eq.~(\ref{LLE}) by a number of authors \citep{matsko_excitation_2012,leo_dynamics_2013,parra-rivas_dynamics_2014} although oscillatory dynamics in similar models had been studied earlier in other contexts \citep{nozaki_chaotic_1985,barashenkov_existence_1996}.

For intermediate values of the detuning (e.g. $\Delta=7$), the bifurcation scenario is like that depicted in Fig.~\ref{breathers1}. The red curve represents the HSS $A_h$, the blue corresponds to the spike state while orange crosses show the maxima and minima of the oscillation. The stable spike encounters a supercritical Hopf bifurcation (H), where it starts to oscillate in amplitude with a single period as shown in the time trace of Fig.~\ref{breathers1}~(1), and in the two-dimensional phase space projection shown in Fig.~\ref{breathers1}~(2), where we also plot $A_h^b$ and the infinite-dimensional saddle corresponding to the unstable spike. Increasing $S$ further, we see that this cycle undergoes a period-doubling bifurcation (PD), starting a route to a very complex scenario via a sequence of oscillatory states characterized by period-2 [see Figs.~\ref{breathers1}~(2),(6)], period-4 [see Figs.~\ref{breathers1}~(3),(7)], and temporal chaos [see Figs.~\ref{breathers1}~(4),(8)]. While increasing $S$, the temporal attractor approaches progressively the saddle spike, leading to a collision that destroys the chaotic attractor, likely in a boundary crisis (BC) \citep{grebogi_crises_1983}. Above BC the only attractor of the system is $A_h^b$. This situation persists until a second BC is reached, where the previous route repeats in reverse order until the system enters into spatiotemporal chaos.

Spatiotemporal chaos was first identified experimentally in fiber cavities \citep{anderson_observations_2016}, and later characterized theoretically in terms of the Lyapunov spectrum and the Yorke-Kaplan dimension \citep{liu_characterization_2017,coulibaly_turbulence-induced_2019}. Furthermore, these studies show that spatiotemporal chaos and the previous dynamical regimes can coexist for the same range of parameters. However, the origin of such dynamics from a bifurcation perspective is not fully understood. The scenario shown in Fig.~\ref{breathers1} summarizes the variety of dynamical behaviors encountered by the system. With further increase in $\Delta$, the scenario remains qualitatively the same although the oscillations are enhanced and the regions of chaotic behavior broaden. Decreasing $\Delta$, however, shifts H towards SN$_1^r$, and the complex oscillatory dynamics gradually fade away, leaving single period oscillations. Finally, H and SN$_1^r$ collide in a codimension-2 bifurcation characterized by three temporal eigenvalues $\lambda_a=0$, and $\lambda_{b,c}=\pm i\omega$, with $\omega>0$ \citep{parra-rivas_dynamics_2014}, commonly known as the Gavrilov-Guckenheimer or Fold-Hopf bifurcation \citep{guckenheimer_nonlinear_1983,gaspard_local_1993}. One of the possible unfoldings of this bifurcation may lead to the appearance of Shilnikov chaos \citep{gaspard_local_1993}, which may be related to the temporal chaos observed here, although confirmation of this scenario requires further investigation.

On top of the dynamics just described, focusing on the single spike LS, one may wonder if LPs coexisting with the previous states below the BD line exhibit a similar dynamical scenario. Indeed, the dynamics of LP-breathers have recently been studied in the context of dispersive optical parametric oscillators, and show very rich and unprecedented behavior \citep{parra-rivas_parametric_2020}. In our current context, however, this question has not been investigated systematically and remains an open problem. 

\begin{figure}[!t]
	\centering\includegraphics[scale=0.8]{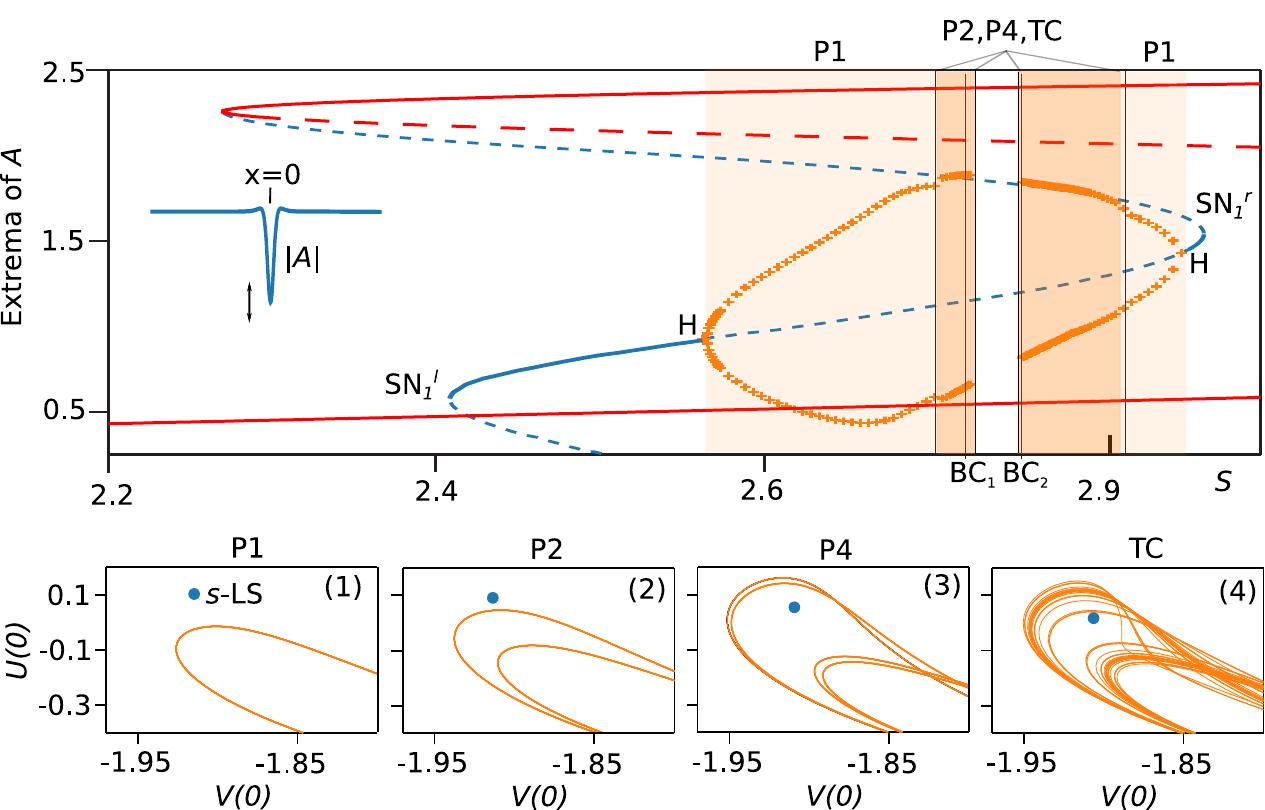}
	\caption{Bifurcation diagram and oscillatory behavior in the normal regime for $\Delta=5.2$. The diagram shows the HSS $\sqrt{I_h}$ in red, the peak values of the spike $|A(0)|$ (blue curve), and the maxima and minima of the oscillation amplitude at $x=0$. The different colored areas correspond to a different types of oscillatory dynamics: period-1 oscillations (P1), period-2 (P2), period-4 (P4) and temporal chaos (TC). The letter H denotes supercritical Hopf bifurcations while BC$_i$ correspond to boundary crises of the chaotic attractor. Panels (1)-(4) show a portion of the two-dimensional phase space projection defined by $U(0)$ and $V(0)$. Adapted from \citet{parra-rivas_dark_2016}.}
	\label{breathers2}
\end{figure}

\subsection{Breathers in the normal regime}
In the normal regime, the oscillatory and chaotic dynamics of LSs have been demonstrated experimentally for the first time in microresonators \citep{bao_observation_2018} and their study is not only restricted to single peak dark states, but also to wider LSs formed by several SOs \citep{parra-rivas_dark_2016,parra-rivas_origin_2016}. Here we briefly discuss some of the main features of the observed dynamics focusing on the single dark spike LS.

Figure~\ref{breathers2} shows the bifurcation scenario for $\Delta=5.2$, where we plot in red the HSS $A_h$ and in blue the spike amplitude at its center ($x=0$). As in the anomalous case, the LS encounters a supercritical Hopf bifurcation (H), where it starts to oscillate in amplitude with a single period, as schematically shown in the inset of Fig.~\ref{breathers2}. In Fig.~\ref{breathers2}~(1) we show a portion of the two-dimensional projection of the cycle in the phase space, together with the projection of the unstable (saddle) LS. Increasing $S$ gradually, the amplitude of the oscillation grows [see Fig.~\ref{breathers2}], and eventually the system undergoes the same dynamical sequence as in the anomalous regime: the single period oscillations undergo a PD bifurcation starting a route to temporal chaos, as depicted in panels (1)-(4) of Fig.~\ref{breathers2}. At some point, the chaotic attractor [see Fig.~\ref{breathers2}~(4)] collides with the saddle, and the system undergoes a boundary crisis BC$_1$ where the attractor is destroyed, opening a parameter window where the systems falls to the static attractor $A_h^t$. This window ends at a second boundary crisis BC$_2$, where temporal chaos reappears again. From this dynamical state, the system undergoes the same bifurcation sequence as just described but in reverse order, until the single period oscillatory state is restored. Increasing $S$ further, the amplitude of the oscillations decreases and eventually the cycle dies at a second supercritical Hopf very close to SN$_1^r$. 

As $\Delta$ increases, the static window between BC$_1$ and BC$_2$ widens as BC$_2$ progressively moves toward H$_2$, and BC$_1$ towards H$_1$. In contrast, decreasing $\Delta$ leads to a fusion of BC$_1$ and BC$_2$, and the disappearance of the static window. With further reduction in $\Delta$, the temporal chaos, period-4 and period-2 cycles gradually fade away, and only the period-1 oscillation remains; for low values of $\Delta$ this oscillation disappears as well \citep{parra-rivas_dark_2016,parra-rivas_origin_2016}. Dark LSs with several SOs undergo a similar behavior to that described here.

\section{Broken spatial reversibility: effect of third order dispersion}\label{sec:6}

In the previous sections we have described the bifurcation structure and main features of different types of localized solutions of Eq.~(\ref{LLE}). This equation describes the dynamics of Kerr dispersive cavities in most practical situations quite well. However, sometimes the modeling of the experimental setup requires the addition of extra terms accounting for a number of physical effects which are not included. Many authors have addressed this issue, and the influence of such terms. In particular, we mention the case of higher-order chromatic dispersion \citep{tlidi_high-order_2010,leo_nonlinear_2013,tlidi_drift_2013,bahloul_temporal_2014,milian_soliton_2014,parra-rivas_third-order_2014}, stimulated Raman scattering (SRS) \citep{milian_solitons_2015,wang_stimulated_2018,clerc_time-delayed_2020-1}, or time-delayed feedback \citep{,panajotov_impact_2016,tlidi_drifting_2017}.

The spatial reversibility of Eq.~(\ref{LLE}) is an essential ingredient not only for the formation of the LSs studied previously, but also for the bifurcation structure undergone by such states. Thus, while high-order terms preserving spatial reversibility (e.g., fourth-order dispersion) lead to similar types of states and bifurcation diagrams \citep{tlidi_high-order_2010}, those breaking it (e.g., third order dispersion or SRS) result in important modifications of the LSs shape, and their dynamics and stability, as well as having strong implications for their bifurcation structure \citep{burke_swift-hohenberg_2009,makrides_predicting_2014}.

In this section, we examine the influence that the loss of spatial reversibility may have on the bifurcation structure of the LSs studied previously, and for this purpose we include the dispersive term $\gamma\partial_x^3A$ accounting for third-order chromatic dispersion (hereafter TOD), in Eq.~(\ref{LLE}):
 \begin{equation}\label{LLE_TOD2}
 \partial_t A= -(1+i\Delta)A+i\nu\partial_x^2 A+\gamma\partial_x^3 A+iA|A|^2+S.
 \end{equation}
As a result of the loss of spatial reversibility, the solutions of Eq.~(\ref{LLE_TOD2}) are asymmetric and drift at constant velocity depending of the control parameters of the system. Steadily drifting LS solutions satisfy the time-independent ordinary differential equation
 \begin{equation}
-(1+i\Delta)A+c\partial_x A+i\nu\partial_x^2 A+\gamma\partial_x^3 A+iA|A|^2+S=0,
 \end{equation}
where the new variable $x$ results from a change in the reference frame $x\rightarrow x-ct$. The resulting solutions can be obtained through path-continuation schemes, with the drift speed $c$ calculated as part of the solution.




\begin{figure}[!t]
	\centering\includegraphics[scale=1]{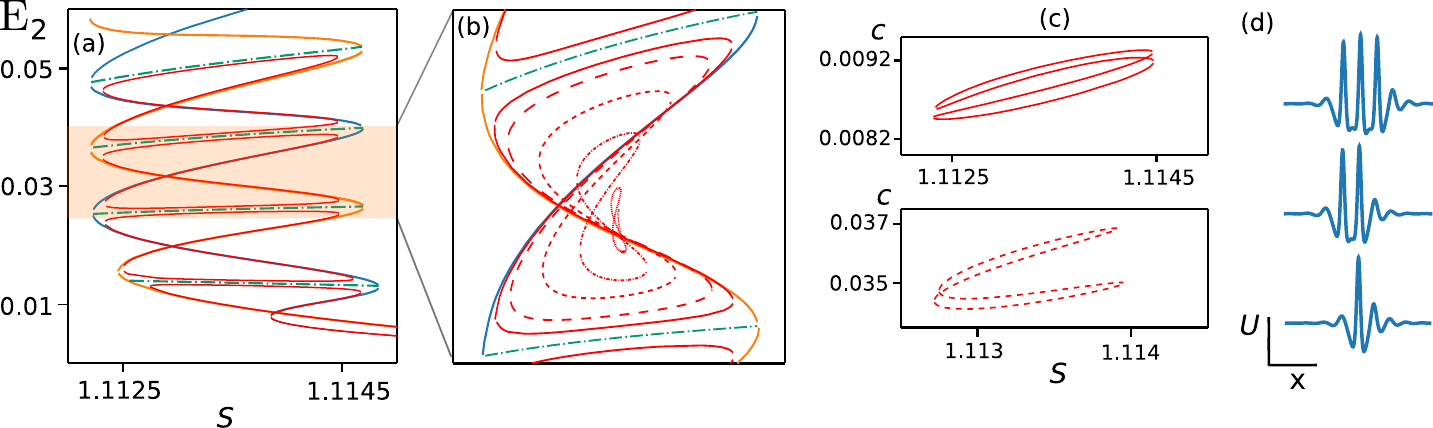}
	\caption{Implications of the loss of spatial reversibility on the snakes-and-ladders structure. Panel (a) shows the formation of a stack of isolas (in red) for $\gamma=0.01$ and the underlying the snakes-and-ladders structure. Panel (b) shows the gradual shrinking of the third isola shown in (a) as $\gamma$ increases, namely $\gamma=0.01,0.02,0.04,0.06,0.076$. Panel (c) shows the speed $c$ of these isola states as a function of $S$ when $\gamma=0.01$ and $\gamma=0.04$. Panel (d) shows three asymmetric LPs when $\gamma=0.04$.}
	\label{Isolas}
\end{figure}

\subsection{Symmetry breaking in the anomalous regime: Isolas of asymmetric states}
The influence of TOD on patterns and LSs dynamics in the anomalous regime have been studied mostly for large detuning, where the typical LSs are spikes. In this context, TOD may lead to the stabilization of oscillatory and chaotic dynamics, and furthermore, to the shrinking of the LSs existence region \citep{milian_soliton_2014,parra-rivas_third-order_2014}.

From a bifurcation perspective, the loss of spatial reversibility is responsible for the destruction of the snakes-and-ladders structure, as first shown in a seminal paper by Burke {\it et al.} in the context of the Swift-Hohenberg equation \citep{burke_swift-hohenberg_2009}.  In the LL equation studied here, the TOD leads to the same destruction for $\Delta<2$, whose main features are summarized in Fig.~\ref{Isolas}. Figure~\ref{Isolas}(a) shows in blue the snakes-and-ladders structure composed of $\Gamma_0$, $\Gamma_\pi$ and the rung states. When $\gamma\neq0$, the pitchfork bifurcations near each SN$_i^{l,r}$ responsible for the rung states become imperfect, leading to the stack of isolas shown in red. Figure~\ref{Isolas}(b) shows a close-up view of the diagram shown in Fig.~\ref{Isolas}(a) around the 3-peak LS branches, where the corresponding isola is shown for several values of $\gamma$. Increasing $\gamma$ leads to the gradual shrinkage of the isolas until they eventually disappear. In the present case this happens for $\gamma\approx0.08$. The speed of the LP along the isola is shown as a function of $S$ in Fig.~\ref{Isolas}(c) for two values of $\gamma$, namely $\gamma=0.01$ (top panel) and $\gamma=0.04$ (bottom panel). Note that an increase in the asymmetry of the LPs due to increasing $\gamma$ results in an increase of their speed. 

The formation of isolas is not the only scenario that one can find in the presence of a reversibility-breaking term \citep{makrides_predicting_2014}. Indeed, for $\Delta>2$, where no snakes-and-ladders structure exists in the absence of TOD, the loss of spatial reversibility leads to a reorganization of $\Gamma_0$ and $\Gamma_\pi$ giving rise to mixed homoclinic snaking \citep{parra-rivas_third-order_2014}. The transition between these two scenarios has been studied in detail in the context of the SH equation \citep{makrides_predicting_2014}. However, in the current context, this point remains an open question.

\begin{figure}[!t]
	\centering\includegraphics[scale=1]{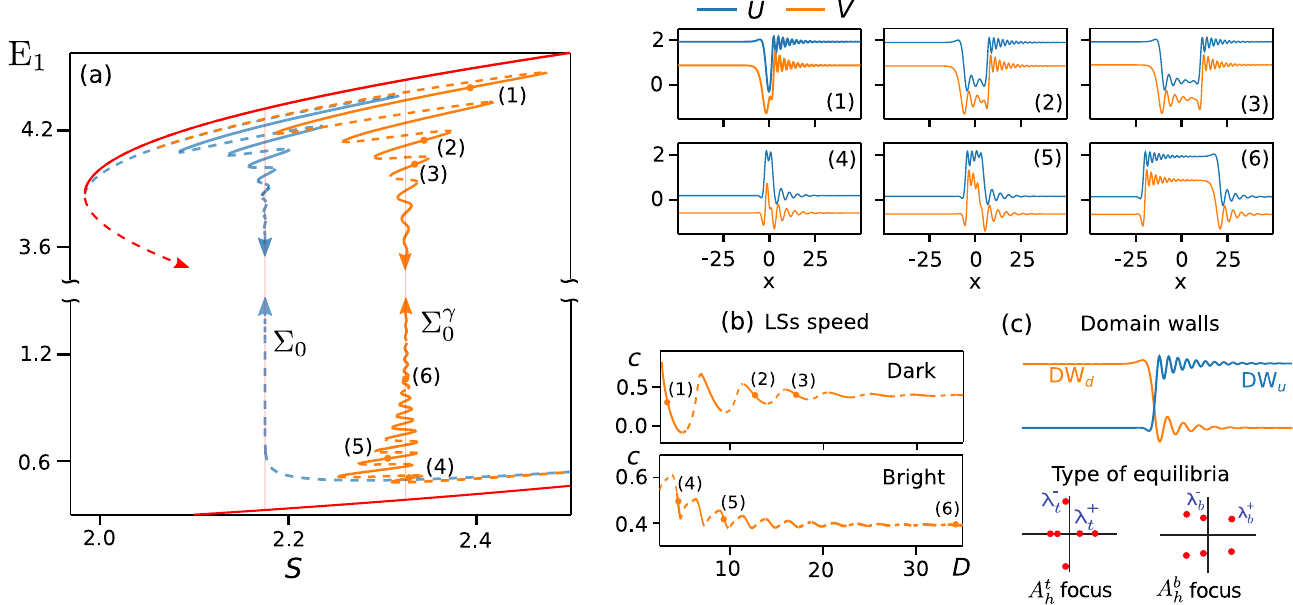}
	\caption{Implications of the loss of spatial reversibility on collapsed snaking. Panel (a) shows in blue the collapsed snaking branch $\Sigma_0$ associated with dark LSs in the absence of TOD. The orange curve $\Sigma_0^\gamma$ correspond to the bifurcation structure computed in the presence of TOD when $\gamma=0.7$. The top part of the diagram shows the change in the collapsed snaking shown in blue, and the labels (1)-(3) correspond to the dark asymmetric states shown on the right panels. The bottom part of the diagram shows the collapsed snaking associated with the asymmetric bright LSs shown in panels (4)-(6). Panel (b) shows the speed $c$ of dark and bright LSs as a function of the LS width $D$. Panel (c) shows the morphology of DW$_d$ and DW$_u$ corresponding to the spatial eigenvalues associated with $A_h^b$ and $A_h^t$. Adapted from \citet{parra-rivas_coexistence_2017}.}
	\label{TOD_collaps}
\end{figure}

\subsection{Symmetry-breaking in the normal regime: Coexistence of dark and bright LSs} 
The influence of TOD on the LSs and their organization in the normal regime has also been addressed in several papers,  both theoretically \citep{he_dynamics_2016,parra-rivas_coexistence_2017,talla_mbe_existence_2017,lobanov_dynamics_2017} and more recently experimentally \citep{li_experimental_2020,anderson_zero-dispersion_2020}. 

From a bifurcation perspective, the loss of spatial reversibility in this regime leads to the coexistence of bright and dark LSs \citep{parra-rivas_coexistence_2017}, and to important modifications of the collapsed snaking morphology as shown in Fig.~\ref{TOD_collaps}. The diagram in blue corresponds to the collapsed snaking branch $\Sigma_0$ shown in Fig.~\ref{collapsed1} in the absence of TOD ($\gamma=0$). The orange curve $\Sigma_0^\gamma$ shows how $\Sigma_0$ changes when spatial reversibility is broken and $\gamma=0.7$. The top part of $\Sigma_0^\gamma$ corresponds to the modified collapsed snaking structure associated with dark LSs. Labels (1)-(3) correspond to the asymmetric dark LSs shown on the right. The bottom part of $\Sigma_0^\gamma$ corresponds to asymmetric bright states such as those shown in Fig.~\ref{TOD_collaps}~(4)-(6). The speed $c$ of the dark and bright LSs along $\Sigma_0^\gamma$ is shown in Fig.~\ref{TOD_collaps}(b) as a function of the LSs width $D$. The speed oscillates along the snaking curves and for large widths it asymptotically reaches a constant value corresponding to a LS like that shown in Fig.~\ref{TOD_collaps}~(6).

The emergence of bright LSs and the modification of collapsed snaking with $\gamma$ can be understood by examining the changes to the interaction and locking of DWs. The breaking of the spatial reversibility is responsible for changing the spatial eigenvalues associated with the equilibria $A_h^b$ and $A_h^t$, which in turn leads to changes in the DWs tails as shown in Fig.~\ref{TOD_collaps}(c). When $\gamma\neq0$, oscillatory tails appear not only around $A_h^b$, but also around $A_h^t$, thereby changing the interaction and the locking of DWs. As a result, stable bright and stable dark LS are both possible, a fact reflected in the change in $\Sigma_0^\gamma$.

Another important effect on the DWs dynamics is that $\gamma$ modifies the position of the Maxwell point $S_M$, and therefore the location of the snaking diagram which shifts to larger values of $S$ as $\gamma$ increases. Moreover, by modifying the DW interaction, TOD is also responsible for an enlargement of the locking regions of the different states, and in consequence of their range of existence in parameter space. The bright states described here, and the collapsed snaking associated with them, have recently been identified in experiments in fiber cavities \citep{li_experimental_2020}.

\section{Discussion and conclussions}\label{sec:7}

It is a general (but useful!) observation that folds of homogeneous solutions of partial differential equations on the real line serve as a source of the spatially modulated and ultimately spatially localized structures found in many (reversible) systems. The reason is simple: in a spatial dynamics description of such systems a fold is associated with a multiplicity two zero eigenvalue. When these become nonzero away from the fold their effect is easily balanced by weak spatial modulation, and it is this balance that leads to the presence of dark solitons near $A_b^t$ and bright solitons near $A_h^b$ in the LL equation. These states are all initially unstable but numerical branch-following techniques show that they typically acquire stability (and hence physical significance) in a process called snaking, following \citet{woods_heteroclinic_1999}. However, this observation is more general and is also responsible for the presence of modulated structures near transcritical bifurcations or indeed near (subcritical) Turing bifurcations, as exemplified by a number of studies of reaction-diffusion equations (see, e.g., \citet{knobloch_stationary_2020} and references therein) or the equations of fluid dynamics \citep{beaume_three-dimensional_2018}. Typically one finds that these localized structures extend between these special points, i.e., that there are (one or more) branches of localized structures connecting these points. This is fundamentally because branches of such states cannot terminate in 'mid-air' or, for physical reasons, extend to infinity. Thus the snaking structures are responsible for the transformation of one spatially extended state of the system into another. In many cases the details of this transformation may be rather complex.

In this paper, we have illustrated these principles using the one-dimensional LL equation, which models a dispersive Kerr optical cavity and is an equation for the intra-cavity electric field envelope in the mean-field approximation. We provided a detailed description of the different types of localized structures arising in this system, unveiling their origin, bifurcation structure, and stability, but never losing sight of the bigger picture. 

The departure point of this work has been the determination of the temporal linear stability properties of the simplest state of the system: the HSS (Sec.~\ref{sec:1.1}). In the anomalous regime, this analysis reveals that a spatially periodic pattern arises from a Turing bifurcation, and that it becomes subcritical in the range $41/30<\Delta<2$, leading to a bistability scenario compatible with a homoclinic snaking structure. In the normal regime, however, the main bistable scenario arises between two HSSs, $A_h^b$ and $A_h^t$, resulting in a collapsed snaking bifurcation structure. Locally, these snaking curves bifurcate from a number of {\it spatial} codimension-one bifurcations of the HSS, including a Hamiltonian-Hopf and a reversible Takens-Bogdanov bifurcation. These bifurcations in turn arise from a codimension-two point known as quadruple-zero (QZ) [occurring at $(\Delta,S)=(2,\sqrt{2})$], which organizes all the behavior in the $(\Delta,S)$-parameter space (Sec.~\ref{sec:1.2}). Near these points, Eq.~(\ref{LLE}) can be reduced to a simpler weakly nonlinear equation (i.e., a {\it normal form}) which retains the essential dynamics of the system. This reduction is performed using multiscale perturbation techniques (Sec.~\ref{sec:2}) and the resulting normal form supports the LS solutions found in Eq.~(\ref{LLE}) near these points. A similar reduction has been also performed around the QZ point (Sec.~\ref{sec:22}). In each case the weakly nonlinear solutions have been tracked away from the bifurcations that gave rise to them using numerical continuation.

In the anomalous regime (Sec.~\ref{sec:3}), the standard snakes-and-ladders structure exists for $\Delta<2$. The LSs corresponding to this scenario are bright localized patterns (Sec.~\ref{sec:3.1}). When $\Delta\rightarrow 2^{-}$, the wavelength of the spatially periodic pattern involved in the formation of the LPs diverges, and at $\Delta=2$ the periodic state undergoes a global homoclinic bifurcation and the different LPs become tame homoclinic orbits (i.e., spikes). This transition destroys the snakes-and-ladders structure and replaces it for $\Delta>2$ by {\it foliated snaking} of spike arrays (Sec.~\ref{sec:3.2}). Furthermore, LPs still form through a heteroclinic tangle below the BD transition, and their solution branches connect to the foliated snaking via a global bifurcation that occurs at the BD point.
 
In the normal regime (Sec.~\ref{sec:4}), the collapsed snaking scenario is present for $\Delta>\sqrt{3}$, and the states associated with it are the dark LSs, consisting in a portion of the low intensity state $A_h^b$ embedded in the high intensity $A_h^t$ background. The formation of this bifurcation structure can be understood through the interaction and locking of DWs \citep{coullet_localized_2002}.

The LL equation (\ref{LLE}) is a non-gradient system and may therefore undergo complex spatio-temporal dynamics such as breathing, temporal chaos and spatiotemporal chaos, in addition to the steady states studied previously. We have shown (Sec.~\ref{sec:5}) that in the anomalous regime bright spikes undergo such dynamics for intermediate values of $\Delta$ as a consequence of a Gavrilov-Guckenheimer bifurcation. In this context, single period oscillatory behavior may undergo a period-doubling route to temporal chaos, and ultimately to spatiotemporal chaos. In the normal regime, a similar scenario is found for dark LSs of different widths. However, in this regime, spatiotemporal chaos is absent.

We have also examined the effects of breaking the spatial reversibility $x\rightarrow-x$ through third-order chromatic dispersion (Sec.~\ref{sec:6}). In the anomalous regime, we have characterized how this symmetry-breaking term destroys the snakes-and-ladders structure leading to a stack of isolas, which eventually fade away as the symmetry-breaking increases. In the normal regime, the collapsed snaking associated with dark LSs persists, but a similar snaking structure emerges as a result of the stabilization of bright LSs. Note that any other term breaking spatial reversibility is expected to lead to a similar bifurcation scenario as has recently been shown when considering the effect of stimulated Raman scattering in the normal regime \citep{parra-rivas_influence_2020}.

There are several issues that have been left out of this work. One of these concerns the interaction of LSs and the formation of bound states. In the anomalous regime, for example, this point has been addressed analytically \citep{vladimirov_effect_2018}, numerically \citep{parra-rivas_interaction_2017} and experimentally \citep{wang_universal_2017}. Another interesting point relates to the effects of higher-order dispersion that preserves spatial reversibility (such as a fourth-order dispersion). In this context, the implications of fourth-order dispersion in the anomalous regime have been analyzed for low values of $\Delta$, where it is responsible for the stabilization of dark LPs and the emergence of new homoclinic snaking \citep{tlidi_high-order_2010}. However, the persistence of homoclinic snaking for larger values of $\Delta$, and a complete understanding of this regime is still lacking. Regarding the normal regime, the impact of such a term on the bifurcation structure of LSs remains an open question. 

Another interesting point relates to the presence of stimulated Raman scattering. This last effect breaks the spatial reversibility of the system, and its implications for the LSs bifurcation scenario and associated dynamics may be similar to those described in Sec.~\ref{sec:6} when dealing with TOD. The dynamics of spike LSs in the presence of the Raman effect have been studied in the anomalous regime by different groups \citep{milian_solitons_2015,wang_stimulated_2018,chen_experimental_2018,sahoo_stability_2019}. However, little is known about its impact on the bifurcation structure associted with spike states or the LPs studied in this paper. In the normal regime, in contrast, the modification of the collapsed snaking structure in the presence of stimulated Raman scattering was recently characterized by \citet{parra-rivas_influence_2020}. 
  
From a mathematical point of view, one may wonder if the previous bifurcation scenarios persist when a two-dimensional version of Eq.~(\ref{LLE}) is considered. Although several works have addressed the study of LSs in the 2D LL equation in the context of diffractive cavities \citep{scroggie_pattern_1994,firth_two-dimensional_1996, firth_dynamical_2002}, the characterization of the bifurcation structure and stability has only focused on single LS \citep{gomila_excitability_2005,gomila_phase-space_2007,gelens_dynamical_2008}, and a complete and systematic characterization is therefore necessary. In this context, the simplest extension of the present work should focus on the radially symmetric structures, described by a nonautonomous one-dimensional problem in the radial coordinate, cf.~\citep{lloyd2009,mccalla2010}. One could even consider a 3D scenario, where LS correspond to the so-called {\it optical bullets}. Although these objects have received some attention in different dissipative systems \citep{jenkins_cavity_2009,veretenov_dissipative_2009,javaloyes_cavity_2016}, their full bifurcation structure remains at present an open problem as well.

\section*{Acknowledgment}
PPR acknowledges support from the Fonds National de la Recherche Scientifique F.R.S.-FNRS (Belgium). This work was supported in part the National Science Foundation under grant DMS-1908891 (EK).


%


\bibliographystyle{imamat}
\bibliography{LLE}

\end{document}